\documentclass[12pt]{iopart}
\usepackage{graphicx}
\usepackage{dcolumn}
\usepackage{rotating}
\usepackage{bm}

\usepackage{iopams}
\def\be{\begin{equation}}
\def\ee{\end{equation}}
\def\ba{\begin{array}}
\def\ea{\end{array}}
\def\bea{\begin{eqnarray}}
\def\eea{\end{eqnarray}}

\begin{document}
\title[\underline{J. Phys. G: Nucl. Part. Phys. \hspace {6cm} Abdul Quddus {\it et al}}]
{Study of hot thermally fissile nuclei using relativistic mean field theory}
\author{\large Abdul Quddus$^1$, K. C. Naik$^2$ and S. K. Patra$^{3,4}$}
\address{$^1$Department of Physics, Aligarh Muslim University, Aligarh-202 002, India\\
$^2$Department of Physics, Siksha 'O' Anusandhan University, Bhubaneswar-751 030, India\\
$^3$Institute of Physics, Sahcivalaya Marg Bhubaneswar-751 005, India\\
$^4$Homi Bhaba National Institute, Training School Complex, Anushakti Nagar, Mumbai 400 085, India}

\date{\today}

\begin{abstract}
We have studied the properties of hot $^{234,236}$U and $^{240}$Pu nuclei in the frame-work of 
relativistic mean field formalism. The recently developed FSUGarnet and IOPB-I parameter sets
are implemented for the first time to deform nuclei at finite temperature. The results are 
compared with the well known NL3 set. The said isotopes are
structurally important because of the thermally fissile nature of $^{233,235}$U and $^{239}$Pu as
these nuclei ($^{234,236}$U and $^{240}$Pu) are formed after the absorption of a thermal neutron,
which undergoes fission. Here, we have evaluated the nuclear properties, such as shell correction energy, 
neutron-skin thickness, quadrupole and hexadecapole deformation parameters and asymmetry energy coefficient for 
these nuclei as a function of temperature.  

\end{abstract}

\pacs{20.10.-k,21.10.Ma,21.10.Pc,21.65.Ef,24.75.+i}
\noindent {Keywords: properties of nuclei, level density, single particle levels, symmetry energy, fission parameters} 
\maketitle

\section{\label{sec:level1} INTRODUCTION}

Out of $\sim$ 300 known stable nuclei in nature, the bottom part of the periodic table, known as actinide series,
encompasses the elements from Z = 89 to 103 which have applications in smoke detectors, gas mantles, as
a fuel in nuclear reactors and in nuclear weapon etc. Among the actinide, Thorium ($Th$) and Uranium ($U$) are 
the most abundant elements in nature with their isotopic fraction: 100\%\ of $^{232}$Th and 0.0054\%\ $^{234}$U, 
0.7204\%\ $^{235}$U and 99.2742\%\ $^{238}$U \cite{nndc}. The isotopes $^{233}$U and $^{239}$Pu are 
synthesized from $^{232}$Th and $^{238}$U, respectively by the bombardment of neutron and subsequent 
$\beta-$decay processes. These actinides are denser than $^{56}$Fe with hardness similar to that of soft steel. 
Apart from their hardness, the naturally occurred $^{235}$U and synthesized $^{233}$U and $^{239}$Pu breakdown 
immediately into fragments with the absorption of slow neutrons (zero energy neutrons or thermal neutrons). The
isotopes $^{233,235}$U and $^{239}$Pu are quite stable in general as long as it is not disturbed by
an almost zero energy external agent, such as a thermal neutron. Hence, these type of isotopes are called as the 
thermally fissile nuclei. These thermally fissile nuclei have a great importance for controlled energy production
in nuclear reactors.

The nuclear fission is one of the most interesting phenomena from the 
time of its discovery by Otto Hahn and F. Strassmann in 1938 \cite{otto}. When a thermally fissile nucleus, 
such as $^{233,235}$U or $^{239}$Pu absorbs  a thermal neutron, it undergoes fission and
releases nuclear energy, which is the main source of energy production in reactor technology.
 After forming the
compound nucleus ($^{234,236}$U and $^{240}$Pu) it oscillates in different modes (quadrupole, hexadecapole) 
of vibrations and finally
reaches to the scission point. In the process, the compound nucleus exhibits various stages including increase
in temperature (T). To understand the fission dynamics, it is important to study the nuclear properties, 
like nuclear excitation energy $E^*$, change in shapes and sizes of nucleus, variation of specific heat $C$, 
effect of shell structure, change of single particle energy and inverse level density parameter. All these 
observables are crucial quantities to understand the fission phenomena and our aim is to analyze these properties 
with temperature.

Recently, the relative mass distribution of thermally fissile nuclei for binary \cite{bharat17} and 
ternary \cite{seinthil17} fission processes are reported. Here, it is shown that the relative yield of 
fission fragments depend very much on the temperature of the system. The level density parameter is also 
influenced a lot with temperature, which is a key quantity in fission study. The neutron-skin thickness 
$\triangle{r_{np}=r_n-r_p}$ ($r_n$ and $r_p$ are the root mean square radii of neutrons and protons distribution)
has a direct correlation with the equation of state (EOS) of nuclear matter
\cite{brown,horo86}. It is to be noted that the neutron star EOS is the main ingredient which is used to predict 
the mass and radius of the star. 
The asymmetry energy coefficient $a_{sym}$ is an important quantity for various nuclear properties, such as
to establish proper boundaries for neutron and proton drip-lines, study of heavy ion collision, 
physics of supernovae explosions and neutron star \cite{barran,prakash,barron85}. Thus, it attracts the attention for the 
analysis of neutron-skin thickness and asymmetry energy coefficient as a function of temperature.

To study the fission process, a large number of models have been proposed
\cite{fong71,chenya88,lest04,fong56,maran09,toke,bord,rama85,swiat,diehl74,wilkins}.
The liquid drop model successfully explains the fission of a nucleus \cite{diehl74,wilkins,pauli,herbert}
and the semi-empirical mass formula is  the simple oldest tool to get a rough 
estimation of the energy released in a binary fission process.
Most of the time, the liquid drop concept is applied to study the fission 
phenomenon, where the shell effect of the nucleus generally  ignored. But, the shell effect plays
a vital role in the stability of the nucleus not only at T=0, but also at finite temperature. This shell 
structure consider to be responsible for the formation of superheavy nuclei in the {\it superheavy island}.
Thus, the microscopic model could be a better frame-work for such type of studies, where the shell structure 
of the nucleus is included automatically. In this aspect the Hartree or Hartree-Fock approach of non-relativistic 
mean field \cite{bender} or relativistic
mean field (RMF) \cite{bogu77} formalisms could be some of the ideal theories. 

The pioneering
work of Vautherin and Brink \cite{vat70}, who has applied the Skyrme interaction in a self-consistent method
for the calculation of ground state properties of finite nuclei opened a new dimension in the quantitative estimation of nuclear properties. Subsequently, the Hartree-Fock and time dependent Hartree-Fock
formalisms \cite{pal} are also implemented to study the properties of fission. Most recently, the microscopic
relativistic mean field approximation, which is another successful theory in nuclear physics is
used for the study of nuclear fission \cite{skp10}. The RMF formalism is not only gaining importance for finite
nuclei, but also quite useful for infinite nuclear matter systems. This theory successfully applied to study 
the gravitational wave strain and the tidal deformability in binary neutron stars merger \cite{bharat95}. 
In the present paper, we have applied the recently developed FSUGarnet \cite{chai15}, and IOPB-I \cite{iopb1} 
models in the framework of temperature dependent effective field theory motivated relativistic mean field (E-RMF) 
formalism, which is the extended version of the standard nonlinear $\sigma-\omega-\rho-$model by including all type
of self and cross-couplings in the RMF Lagrangian. For comparison, the widely used NL3 force \cite{lala97} is also 
applied in the calculations. Since, thermally fissile nuclei $^{233}$U, 
$^{235}$U and $^{239}$Pu undergo fission through the $^{234}$U, $^{236}$U 
and $^{240}$Pu respectively, as mentioned earlier, after absorbing a thermal 
neutron. Thus, we have studied the properties of hot $^{234}$U, $^{236}$U and 
$^{240}$Pb nuclei. Along with these nuclei, we have taken $^{208}$Pb as a 
representative case to examine our calculation for spherical nuclei.

The paper is organized as follows: In Section II, the finite temperature relativistic mean field theory is 
presented briefly.  In this section, the equation of motion of the nucleon and meson fields are obtained 
from the relativistic mean field Lagrangian.  The temperature dependent of the equations are adopted through 
the occupation number of protons and neutrons as it is developed in Refs. \cite{bharat17,seinthil17}. The 
results are discussed in Section III and compared with various models, wherever necessary. The summary 
and concluding remarks are given in Section IV.

\section{THEORETICAL FRAMEWORK}

\subsection{Effective field theory motivated relativistic mean field formalism}\label{sec:bcs}

From last five decades the relativistic mean field (RMF) formalism is one of the most successful and widely used 
theory for both finite nuclei and infinite nuclear matter systems including the study of neutron star (NS). It is 
nothing but the relativistic generalization of the non relativistic Hartree or Hartree-Fock Bogoliubov theory with the 
advantages that it takes into account the spin orbit interaction automatically and works better in high density 
region. In this theory, nucleons are considered to oscillate independently in a harmonic oscillator motion in the 
mean fields generated by the exchange of mesons and photons. The nucleons are interacted with each other through 
the exchange of isoscalar-scalar $\sigma-$, isoscalar-vector $\omega-$ and
isovector-vector $\rho-$mesons. The $\sigma$, $\omega$ and photon fields are taken into account in the original
RMF formalism, known as the linear $\sigma-\omega$ model. In this approximation, the model predicts incompressibility 
$K_{\infty}$ of nuclear matter $\sim$ 550 MeV which is far away from the experimental value $210 \pm 30$ MeV
\cite{blaizot}.  To rectify this limitation, Boguta and Bodmer included the self couplings of $\sigma$ meson in 
the Lagrangian and hence, known as the non-linear $\sigma -\omega$ model \cite{bogu77} which reduces the 
incompressibility $K_{\infty}$ of nuclear matter to a considerable range of $\sim 200-300$ MeV. It reproduces the 
nuclear bulk properties like binding energy BE, root mean square (rms) charge radius $R_{ch}$, neutron-skin thickness
$\triangle{r_{np}}$ and quadrupole deformation parameter $\beta_2$ etc. remarkably well not only for the 
$\beta-$stable nuclei, but also for exotic nuclei which are far away from the $\beta-$stability valley.
After the success of this $\sigma-\omega-\rho-$model, a large number of force parameters have been proposed,
which take into account various other mediating mesons and their self and cross couplings. These higher order
couplings  are quite important, because each and every coupling in the E-RMF Lagrangian has its own effect
to explain various physical phenomena in different environments starting from very low to high density domains.
As a result, parameter sets like FSUGarnet \cite{chai15}, IOPB-I \cite{iopb1}, 
G1 \cite{furnstahl97}, G2 \cite{furnstahl97} and G3 \cite{bharatnpa} have 
been evolved with time.

In this section, we briefly outline the effective field theory motivated relativistic mean field (E-RMF) 
theory \cite{furnstahl97}. The model is used by fitting the coupling constants and mass of the $\sigma-$meson
to reproduce the known nuclear ground state properties of some spherical nuclei as well as the nuclear matter
properties at saturation. In principle, the  E-RMF Lagrangian has an infinite number of terms with all type of self
and cross couplings. Thus, it is necessary to develop a truncation scheme to handle numerically for the
calculations of finite and infinite nuclear matter properties. 
The meson fields included in the Lagrangian are smaller than the
mass of nucleon. Their ratios are used as a truncation scheme as it
is done in Refs. \cite{furnstahl97,muller96,serot97,estal01}. This means $\Phi/M$, $W/M$ and $R/M$
are the expansion parameters. The constraint of naturalness is also introduced in the truncation
scheme to avoid ambiguities in the expansion. In other words, the coupling constants written with a
suitable dimensionless form should be $\sim 1$. Imposing these conditions, one can then estimate the
contributions coming from different terms of the Lagrangian by counting powers in the expansion up to
a certain order of accuracy and the coupling constants should not be truncated arbitrarily. 
It is shown that the Lagrangian up to fourth order of dimension is a good approximation to predict 
finite nuclei and nuclear matter observables up to considerable satisfaction \cite{furnstahl97,muller96,serot97,estal01}.
Thus, in the present calculations, we have considered the contribution of the terms in the E-RMF Lagrangian up to
$4^{th}$ order of expansion. The nucleon-meson E-RMF Lagrangian density 
with  $\sigma-$, $\omega-$, $\rho-$mesons and photon $A^{\mu}$ fields is given as
\cite{chai15,iopb1}:
\begin{eqnarray}
{\cal L}({r_{\perp},z}) & = & \bar\varphi({r_{\perp},z}) \left(i\gamma^\mu\partial\mu -M + g_s \sigma - g_\omega \gamma^\mu \omega_\mu - 
g_\rho \gamma^\mu \tau \vec{\rho}_\mu 
-e \gamma^\mu \frac{1+\tau_3}{2} A_\mu \right) \cdot
\nonumber \\[3mm]
& &
\varphi({r_{\perp},z})
+ \frac{1}{2}\left(\partial^\mu\sigma\partial_\mu\sigma-m_s^2\sigma^2 \right)
-\frac{1}{4}V^{\mu\upsilon}V_{\mu\upsilon} + \frac{1}{2}m_\omega^2 \omega^\mu \omega_\mu
\nonumber \\[3mm]
& &
-\frac{1}{4} \vec{R}^{\mu\upsilon} \vec{R}_{\mu\upsilon} 
+ \frac{1}{2}m_\rho^2\vec{\rho}^\mu\vec{\rho}_\mu-\frac{1}{4}F^{\mu\upsilon} F_{\mu\upsilon}
- m_s^2 \sigma^2 \left(\frac{k_3}{3!}\frac{g_s\sigma}{M} + \frac{k_4}{4!}\frac{g_s^2\sigma^2}{M^2} \right ) 
\nonumber \\[3mm]
& &
+\frac{1}{4!} \zeta_0 g_\omega^2 (\omega^\mu \omega_\mu)^2
+ \Lambda_\omega g_\omega^2 g_\rho^2(\omega^\mu \omega_\mu)(\vec{\rho}^\mu \vec{\rho}_\mu)
\;,
\end{eqnarray}
with 
\begin{eqnarray}
V^{\mu\upsilon} & = & \partial^\mu \omega^\upsilon -\partial^\upsilon \omega^\mu
\;,
\end{eqnarray} 
\begin{eqnarray}
\vec{R}^{\mu\upsilon} & = & \partial^\mu \vec{\rho}^\upsilon -\partial^\upsilon\vec{\rho}^\mu - g_{\rho} (\vec{\rho}^\mu \times  \vec{\rho}^\upsilon )
\;,
\end{eqnarray}
\begin{eqnarray}
F^{\mu\upsilon} & = & \partial^\mu A^\upsilon -\partial^\upsilon A^\mu
\;.
\end{eqnarray}

Here $\sigma$, $\omega$ and $\rho$ are the mesonic fields having masses $m_s$, $m_\omega$ and $m_\rho$ 
with coupling constants $g_s$, $g_\omega$ and $g_\rho$ for $\sigma-$, $\omega-$ and $\rho-$mesons, respectively. 
$A^\mu$ is the photon field which is responsible for electromagnetic interaction with coupling strength 
$\frac{e^2}{4\pi}$. By using variational principle and applying mean field approximations, the equations 
of motion for the nucleon and boson fields are obtained. Redefining fields as $\Phi = g_s\sigma $, 
$W = g_\omega \omega^0$, $R = g_\rho \vec{\rho}^0$ and $A = eA^0 $, the Dirac equation corresponds to 
the above Lagrangian density is
\begin{eqnarray}
\Bigg\{-i \mbox{\boldmath$\alpha$} \!\cdot\! \mbox{\boldmath$\nabla$}
& + & \beta [M - \Phi(r_{\perp},z)] + W(r_{\perp},z)
  +\frac{1}{2} \tau_3 R(r_{\perp},z) 
\nonumber \\[3mm] 
& &
+ \frac{1 +\tau_3}{2}A(r_{\perp},z)
     \bigg\}  
\varphi_\alpha (r_{\perp},z) =
     \varepsilon_\alpha \, \varphi_\alpha (r_{\perp},z) \,.
     \label{eq5}
     \end{eqnarray}
The mean field equations for $\Phi$, $W$, $R$ and $A$ are given as
\begin{eqnarray}
-\Delta \Phi(r_{\perp},z) + m_s^2 \Phi(r_{\perp},z)  & = &
g_s^2 \rho_s(r_{\perp},z)
-{m_s^2\over M}\Phi^2 (r_{\perp},z)
\left({\kappa_3\over 2}+{\kappa_4\over 3!}{\Phi(r_{\perp},z)\over M}
\right )
 \,,
 \label{eq6}  \\[3mm]
-\Delta W(r_{\perp},z) +  m_\omega^2 W(r_{\perp},z)  & = &
g_\omega^2  \rho(r_{\perp},z) 
-{1\over 3!}\zeta_0 W^3(r_{\perp},z)
-2\;\Lambda_{\omega} g_\omega^2 {R^{2}(r_{\perp},z)} W(r_{\perp},z) \,,
\label{eq7}  \\[3mm]
-\Delta R(r_{\perp},z) +  m_{\rho}^2 R(r_{\perp},z)  & = &
{1 \over 2 }g_{\rho}^2 \rho_{3}(r_{\perp},z) 
-2\;\Lambda_{\omega} g_{\rho}^2 R(r_{\perp},z) {W^{2}(r_{\perp},z)} \,,
\label{eq8}  \\[3mm]
-\Delta A(r_{\perp},z)   & = &
e^2 \rho_{\rm p}(r_{\perp},z)    \,.
\label{eq9}
\end{eqnarray}
The baryon, scalar, isovector and proton densities used in the above equations are defined as
\begin{eqnarray}
\rho(r_{\perp},z) & = &
\sum_\alpha n_{\alpha} \varphi_\alpha^\dagger(r_{\perp},z) \varphi_\alpha(r_{\perp},z) 
\nonumber\\
        &=&\rho_{p}(r_{\perp},z)+\rho_{n}(r_{\perp},z) \,,
\label{den_nuc}
\end{eqnarray}
\begin{eqnarray}
\rho_s(r_{\perp},z) & = &
\sum_\alpha n_{\alpha} \varphi_\alpha^\dagger(r_{\perp},z) \beta \varphi_\alpha(r_{\perp},z)
\nonumber \\
        &=&\rho_{s p}(r_{\perp},z) +  \rho_{s n}(r_{\perp},z) \,,
\label{scalar_density}
\end{eqnarray}
\begin{eqnarray}
\rho_3 (r_{\perp},z) & = &
\sum_\alpha n_{\alpha} \varphi_\alpha^\dagger(r_{\perp},z) \tau_3 \varphi_\alpha(r_{\perp},z) \nonumber \\
&=& \rho_{p} (r_{\perp},z) -  \rho_{n} (r_{\perp},z) \,,
\label{isovect_density}
\end{eqnarray}
\begin{eqnarray}
\rho_{\rm p}(r_{\perp},z) & = &
\sum_\alpha n_{\alpha} \varphi_\alpha^\dagger(r_{\perp},z) \left (\frac{1 +\tau_3}{2}
\right)  \varphi_\alpha(r_{\perp},z).
\label{charge_density}
\end{eqnarray}
Where $\beta$ and $\tau_3$ have their usual meanings. We have taken summation 
 over $\alpha$ where $\alpha$ stands for all nucleon. The factor $n_{\alpha}$, 
used in the density expressions is nothing but occupation probability which is 
described in the next sub-section. The effective mass of 
nucleon due to its motion in the mean field potential is given as
$M^{\ast}=M- \Phi (r_{\perp},z)$ and the 
vector potential is
$V(r_{\perp},z)=W(r_{\perp},z)+\frac{1}{2} \tau_{3}R(r_{\perp},z)
+\frac{(1+\tau_3)}{2}A(r_{\perp},z).$
The energy densities for nucleonic and mesonic fields corresponding to the Lagrangian density are
\begin{eqnarray}
{\cal E}_{nucl.}({r_{\perp},z}) & = &  \sum_\alpha \varphi_\alpha^\dagger({r_{\perp},z})
\Bigg\{ -i \mbox{\boldmath$\alpha$} \!\cdot\! \mbox{\boldmath$\nabla$}
+ \beta \left[M - \Phi (r_{\perp},z) \right] + W({r_{\perp},z})
\nonumber \\[3mm]
& &
+ \frac{1}{2}\tau_3 R({r_{\perp},z})
+ \frac{1+\tau_3}{2} A ({r_{\perp},z})\bigg\}\varphi_\alpha (r_{\perp},z)
\;,
\label{eq11}
\end{eqnarray}
and
\begin{eqnarray}
{\cal E}_{mes.}({r_{\perp},z}) & = & \frac{1}{2g_s^2}\left [\left(\mbox{\boldmath $\nabla$}\Phi({r_{\perp},z})\right)^2
+ m_s^2 \Phi^2({r_{\perp},z})\right]
  + \left (  \frac{\kappa_3}{3!}\frac{\Phi({r_{\perp},z})}{M}
  + \frac{\kappa_4}{4!}\frac{\Phi^2({r_{\perp},z})}{M^2}\right )
\nonumber \\[3mm]
& &
\cdot   \frac{m_s^2}{g_s^2} \Phi^2({r_{\perp},z})
- \frac{1}{2g_\omega^2}\left[\left(\mbox{\boldmath $\nabla$}W({r_{\perp},z})\right)^2+{m_\omega^2} W^2 ({r_{\perp},z})\right]
\nonumber \\[3mm]
& &
- \frac{1}{2g_\rho^2}\left[\left(\mbox{\boldmath $\nabla$}R({r_{\perp},z})\right)^2+m_\rho^2 R^2 ({r_{\perp},z})\right]
-  \frac{\zeta_0}{4!} \frac{1}{ g_\omega^2 } W^4 ({r_{\perp},z})
\nonumber \\[3mm]
& & 
-\Lambda_{\omega}R^{2}(r_{\perp},z)\!\cdot\! W^{2}(r_{\perp},z) 
- \frac{1}{2e^2}\left[\left(\mbox{\boldmath $\nabla$}A({r_{\perp},z})\right)^2 \right] 
\;.
\end{eqnarray}
To solve the set of coupled differential equations (5-9) 
we expand the Boson and Fermion fields in an axially deformed
harmonic oscillator basis with $\beta_0$ as the initial deformation.
The set of equations are solved self iteratively till the convergence is 
achieved.
The center of mass correction is subtracted within the non-relativistic approximation \cite{negele}. The calculation
is extended to finite temperature T through the occupation number $n_{\alpha}$ in the BCS pairing formalism.
The quadrupole deformation parameter
$\beta_2$ is estimated from the resulting quadrupole moments of the
protons and neutrons as $Q = Q_n + Q_p = \sqrt{\frac{16\pi}5} (\frac3{4\pi} AR^2\beta_2)$, (where $R=1.2A^{1/3}$fm ). 
The total energy at finite temperature T is given by \cite{blunden87,reinhard89,gam90},
\begin{eqnarray}
E(T) = \sum_{\alpha} \epsilon_{\alpha} n_{\alpha} + E_{mes.} +E_{pair} + E_{c.m.} - AM,
\end{eqnarray}
with
\begin{equation}
E_{mes.}  =  \int d^3r {\cal E}_{mes.}({r_{\perp},z}) ,
\end{equation}
\vspace{-3mm}
\begin{eqnarray}
E_{pair} = - \triangle\sum_{\alpha>0}u_{\alpha}v_{\alpha} = -\frac{\triangle^2}{G},
\end{eqnarray}
\vspace{-3mm}
\begin{equation}
E_{c.m.}= -\frac{3}{4}\times 41A^{-1/3}.
\end{equation}
Here, $\epsilon_{\alpha}$ is the single particle energy, $n_{\alpha}$ is the occupation 
probability and $E_{pair}$ is the pairing energy obtained from the BCS formalism. The $u_{\alpha}^2$ and
$v_{\alpha}^2$ are the probabilities of unoccupied and occupied states, respectively.

\subsection{Pairing and temperature dependent E-RMF formalism}\label{sec:bcs}

There are experimental evidences that even-even nuclei are more stable than even-odd or odd-odd isotopes.
Thus, pairing correlation plays a distinct role in an open shell nuclei. The total binding energy of open shell 
nuclei deviates slightly from the experimental value when pairing correlation is not considered. To explain
this effect, Aage Bohr, Ben Mottelson and Pines suggested BCS pairing in nuclei \cite{bohr} just after the
formulation of BCS theory for electrons in metals.  The BCS pairing in nuclei is analogous to the pairing 
of electrons (Cooper pair) in super-conductors. It is used to explain energy gap in single particle spectrum.
The detail formalism is given in Refs. \cite{bharat17,seinthil17}, but for completeness, we are briefly highlighting 
some essential part of the formalism.  The BCS pairing state is defined as 
\begin{equation}
\mid \Psi_0 >^{BCS} = \prod_{j,m>0}(u_j + 
v_j\varphi_{j,m}^{\dagger}\varphi_{j,-m}^{\dagger})\mid 0>,
\end{equation}
where j and m are the quantum numbers of the state.
In the mean field formalism, the violation of particle number is seen due to the pairing correlation, i.e.,
the appearance of terms like $\varphi^{\dagger}\varphi^{\dagger}$ or $\varphi\varphi$, which are responsible for pairing
correlations. Thus, we neglect such type of interaction at the RMF level and taking externally the pairing 
effect through the constant gap BCS pairing. 
The  pairing interaction energy in terms of occupation
probabilities $v_{\alpha}^2$ and $u_{\alpha}^2=1-v_{\alpha}^2$ (where $\alpha$ stands for nucleon) is written
as~\cite{patra93,pres82}:
\begin{equation}
E_{pair}=-G\left[\sum_{\alpha>0}u_{\alpha}v_{\alpha}\right]^2,
\end{equation}
with $G$ is the pairing force constant.
The variational approach with respect to the occupation number $v_{\alpha}^2$ gives the BCS equation
\cite{pres82}:
\begin{equation}
2\epsilon_{\alpha}u_{\alpha}v_{\alpha}-\triangle(u_{\alpha}^2-v_{\alpha}^2)=0,
\label{eqn:bcs}
\end{equation}
with the pairing gap $\triangle=G\sum_{\alpha>0}u_{\alpha}v_{\alpha}$. The pairing gap ($\triangle$) of proton 
and neutron is taken from the empirical formula \cite{gam90,va73}:
\begin{equation}
\triangle = 12 \times A^{-1/2}.
\end{equation}
The temperature introduced in the partial occupancies through the BCS approximation is given by,
\begin{equation}
n_{\alpha}=v_{\alpha}^2=\frac{1}{2}\left[1-\frac{\epsilon_{\alpha}-\lambda}{\tilde{\epsilon_{\alpha}}}[1-2 f(\tilde{\epsilon_{\alpha}},T)]\right],
\end{equation}
with
\begin{eqnarray}
f(\tilde{\epsilon_{\alpha}},T) = \frac{1}{(1+exp[{\tilde{\epsilon_{\alpha}}/T}])} ,  &  \nonumber \\[3mm]
\tilde{\epsilon_{\alpha}} = \sqrt{(\epsilon_{\alpha}-\lambda)^2+\triangle^2}.&
\end{eqnarray}
The function $f(\tilde{\epsilon_{\alpha}},T)$ represents the Fermi Dirac distribution for quasi particle 
energy $\tilde{\epsilon_{\alpha}}$. The chemical potential $\lambda_p (\lambda_n)$
for protons (neutrons) is obtained from the constraints of particle number equations
\begin{eqnarray}
\sum_{\alpha} n_{\alpha}^{Z}  = Z, \nonumber \\
\sum_{\alpha} n_ {\alpha}^{N} =  N.
\end{eqnarray}
The sum is taken over all the protons and neutrons states. The entropy is obtained by,
\begin{equation}
S = - \sum_{\alpha} \left[n_{\alpha}\, lnn_{\alpha} + (1 - n_{\alpha})\, ln (1- n_{\alpha})\right].
\end{equation}
The total energy and the gap parameter are obtained by minimizing the free energy,
\begin{equation}
F = E - TS.
\end{equation}
In constant pairing gap calculations, for a particular value of pairing gap $\triangle$
and force constant $G$, the pairing energy $E_{pair}$ diverges, if
it is extended to an infinite configuration space. In fact, in all
realistic calculations with finite range forces, $\triangle$ is not
constant, but decreases with large angular momenta states above the Fermi
surface. Therefore, a pairing window in all the equations are extended
up-to the level $|\epsilon_{\alpha}-\lambda|\leq 2(41A^{-1/3})$ as a function
of the single particle energy. The factor 2 has been determined so as
to reproduce the pairing correlation energy for neutrons in $^{118}$Sn
using Gogny force \cite{gam90,patra93,dech80}.

\section{RESULTS AND DISCUSSIONS}

The detail results of our calculations are presented in Table 2 and Figures $1-11$. Here, we discuss the binding
energy, nuclear radii, quadrupole and hexadecapole deformation parameters, specific heat, shell correction energy, inverse
level density parameter, two neutron separation energy and asymmetry energy coefficient as a function of temperature T.
First of all we will explain our motivation for the choice of parameter sets used and subsequently we will discuss the results of 
our calculations. 

\begin{table}
\caption{The parameter sets FSUGarnet \cite{chai15}, IOPB-I \cite{iopb1} and NL3 \cite{lala97}
used in the calculations are  listed. The nucleon mass $M$ taken as 939.0 MeV.  All the coupling 
constants are dimensionless, except $k_3$ which is in fm$^{-1}$. The lower panel of the table 
shows the nuclear matter properties of the models \cite{iopb1}.}
\scalebox{1.1}{
\begin{tabular}{cccccccccc}
\hline
\hline
\multicolumn{1}{c}{}
&\multicolumn{1}{c}{NL3}
&\multicolumn{1}{c}{FSUGarnet}
&\multicolumn{1}{c}{IOPB-I}\\
\hline
$m_{s}/M$  &  0.541  &  0.529& 0.533  \\
$m_{\omega}/M$  &  0.833  & 0.833 &0.833  \\
$m_{\rho}/M$  &  0.812 & 0.812 &  0.812  \\
$g_{s}/4 \pi$  &  0.813  &  0.837 &0.827 \\
$g_{\omega}/4 \pi$  &  1.024  & 1.091 &1.062 \\
$g_{\rho}/4 \pi$  &  0.712  & 1.105 &0.885  \\
$k_{3} $   &  1.465  & 1.368 &1.496 \\
$k_{4}$  &  -5.688  &  -1.397 &-2.932  \\
$\zeta_{0}$  &  0.0  &4.410  &3.103  \\
$\Lambda_{\omega}$  &  0.0  &0.043& 0.024   \\
\hline
$\rho$ (fm$^{-3})$ &  0.148  &  0.153&0.149  \\
$\mathcal{E}_{0}$(MeV)  &  -16.29  & -16.23 &-16.10  \\
$M^{*}/M$  &  0.595 & 0.578 &  0.593  \\
$J$(MeV)   & 37.43  &  30.95&  33.30  \\
$L$(MeV)  &  118.65  &  51.04 &63.58 \\
$K_{sym}$(MeV)  &  101.34  & 59.36 & -37.09 \\
$Q_{sym}$(MeV)  &  177.90  & 130.93&  862.70  \\
$K_{\infty}$(MeV)  & 271.38  &  229.5&  222.65 \\
$Q_{0} $(Mev)   &  211.94  & 15.76& -101.37 \\
$K_{\tau}$(MeV)  &  -703.23  &  -250.41&-389.46  \\
$K_{asy}$(MeV)  & -610.56  & -246.89&  -418.58 \\
$K_{sat2}$(MeV)  & -703.23  &-250.41&  -389.46 \\

\hline
\hline
\end{tabular}}
\label{table1}
\end{table}

\begin{table}
\caption{The ground state binding energy per nucleon, quadrupole deformation parameter and the charge radius 
for $^{208}$Pb, $^{234,236}$U and $^{240}$Pu corresponding to FSUGarnet 
\cite{chai15}, IOPB-I \cite{iopb1} and NL3 \cite{lala97} 
sets are compared with the experimental data \cite{nndc,Angeli2013}.}
\scalebox{1.1}{
\begin{tabular}{cccccccccc}
\hline
\hline
\multicolumn{1}{c}{Nucleus}
&\multicolumn{1}{c}{Observables}
&\multicolumn{1}{c}{FSUGarnet}
&\multicolumn{1}{c}{IOPB-I}
&\multicolumn{1}{c}{NL3}
&\multicolumn{1}{c}{Exp.}\\
\hline
$^{208}$Pb  & B/A(MeV)  & 7.88 & 7.88  &  7.88& 7.87  \\
 & $R_{\rm ch}$ (fm) &5.55  &5.58  &5.52& 5.50 \\
 &$\beta_2$ &  0.00& 0.00 &  0.00& 0.00 \\ 
$^{234}$U  & B/A(MeV)  & 7.60 & 7.61  & 7.60 & 7.60  \\
 &$R_{\rm ch}$ (fm) & 5.84& 5.88  & 5.84 & 5.83  \\
 &$\beta_2$ & 0.20 & 0.20  & 0.24 & 0.27  \\ 
$^{236}$U  & B/A(MeV)  & 7.57& 7.59  & 7.58 & 7.59  \\
 &$R_{\rm ch}$ (fm) & 5.86 & 5.90  &  5.86 & 5.84  \\
 &$\beta_2$ & 0.22 & 0.22  &  0.25 & 0.27  \\ 
$^{240}$Pu  & B/A(MeV)  & 7.56 & 7.57  & 7.55  & 7.56  \\
 &$R_{ch}$ (fm) & 5.91 & 5.95  &  5.90 & 5.87  \\
 &$\beta_2$ & 0.24 & 0.25 & 0.27 & 0.29  \\ 
\hline
\hline
\end{tabular}}
\label{table2}
\end{table}

\subsection{Parameter chosen}\label{sec:para}

There are large number of parameter sets $\sim 265$ available in the literature \cite{iopb1,bharatnpa,dutra12}. All the forces are 
designed with an aim to explain certain nuclear phenomena either in normal or in extreme conditions.  
In a relativistic mean field Lagrangian, every coupling term has its own effect to explain some
physical quantities. For example, the self-coupling terms in the $\sigma-$meson takes care of the
3-body interaction which helps to explain the Coester band problem and the incompressibility coefficient 
$K_{\infty}$ of nuclear matter at saturation \cite{bogu77,fujita,pie}. In the absence of these non-linear $\sigma-$couplings, 
the earlier force parameters predict a large value of $K_{\infty}\sim 540$ MeV \cite{serot86,wal74}. The non-linear
term of the isoscalar vector meson plays a crucial role to soften the nuclear equation of state (EOS) \cite{toki}. 
Adjusting this coupling constant $\zeta_0$, one can reproduce the experimental data of the sub-saturation density \cite{bharatnpa}. The
finite nuclear system is in the region of sub-saturation density and this coupling could be important
to describe the phenomena of finite nuclei.  

For a quite some time, the cross coupling of the isovector-vector $\rho-$meson and the 
isoscalar-vector $\omega-$meson is ignored in the calculations. Even the effective field theory motivated
relativistic mean field (E-RMF) Lagrangian \cite{furnstahl97} does not include this term in its original
formalism. For the first time, Todd-Rutel and Piekarewicz \cite{pika05} realized the effect of this coupling
in the correlation of neutron-skin thickness and the radius of neutron star.  This coupling constant 
$\Lambda_{\omega}$ influences the neutron distribution radius $r_n$ without affecting much other properties
like proton distribution radius $r_p$ or binding energy of finite nucleus. The RMF parameter set without 
$\Lambda_{\omega}$ coupling predicts a larger incompressibility coefficient $K_{\infty}$ than the non-relativistic
Skyrme/Gogny interactions or empirical data. However, this value of $K_{\infty}$ agree with such predictions
when the $\Lambda_{\omega}$ coupling present in the Lagrangian. Thus, the parameter $\Lambda_{\omega}$
can be used as a bridge between the non-relativistic and relativistic mean field models. Although, the 
contribution of this coupling is marginal for the calculation of bulk properties of finite
nuclei, the inclusion of $\Lambda_{\omega}$ in the E-RMF formalism  is important
for its softening nature to nuclear equation of states. The inclusion of non-linear term of $\omega$ field and 
cross coupling of vector fields ($\Lambda_{\omega}$) reproduce 
experimental values of GMR and IVGDR well comparative to those of NL3, and hence, they are needed for reproducing 
a few nuclear collective modes \cite{pika05}. These two terms also soften both EOS of symmetric nuclear matter 
and symmetry energy.

In the present paper, the results are obtained from three different RMF sets, 
namely FSUGarnet
\cite{chai15}, IOPB-I \cite{iopb1} and NL3 \cite{lala97}. The NL3 is the oldest among them and one of the most successful force
for finite nuclei all over the mass table. It produces excellent results for binding energy, charge
radius and quadrupole deformation parameter not only for $\beta-$stability nuclei, but also for nuclei away
from the valley of stability. On the other hand the FSUGarnet is a recent parameter set \cite{chai15}. It is
seen in Ref. \cite{iopb1} that this set reproduces the neutron-skin thickness $\triangle{r_{np}}=r_n-r_p$ with the
recent data up to a satisfactory level along with other bulk properties. The IOPB-I is the latest in this series and
reproduces the results with an excellent agreement with the data. It is to be noted that the FSUGarnet reproduces 
the neutron star mass in the lower limit, i.e., $M = 2.06 M_{\odot}$ and the IOPB-I gives the upper limit of 
neutron star mass $M = 2.15 M_{\odot}$ \cite{iopb1}. These FSUGarnet and IOPB-I parameters have the additional 
non linear term of isoscalar vector meson and cross coupling of vector mesons $-\omega, -\rho$ over NL3 set.
To our knowledge, for the first time the
FSUGarnet and IOPB-I are used for the calculations of deform nuclei. Also, for the first time these two
sets are applied to finite temperature calculations for nuclei. The values of the parameters and their nuclear
matter properties are depicted in Table 1. 

First of all, we want to compare our calculated results with the experimental data. The results of our calculations 
for binding energy per particle B/A, quadrupole deformation parameter $\beta_2$ and charge radius $R_{ch}$ for
$^{208}$Pb, $^{234,236}$U and $^{240}$Pu using FSUGarnet, IOPB-I and NL3 are tabulated in Table 2 at zero 
temperature. The experimental data are also given for comparison. From the table, it is clear that all the
parameter sets reproduce the results remarkably well. A further inspection of the table indicates that some time,
the binding energy predicted by IOPB-I overestimates the data. However, the deformation parameter $\beta_2$
is slightly smaller and the charge radius $R_{ch}$ is slightly larger compared to the experimental measurements.  In general, all 
the three observables are in an excellent agreement with the experimental observations and we can use the models
for further predictions at different conditions, such as at finite temperature.
Now we want to discuss some important nuclear properties at 
finite temperature with these three successful models in the following
sub-sections.
\begin{figure}
        \hspace{0.8cm}
        \includegraphics[width=0.9\columnwidth,clip=true]{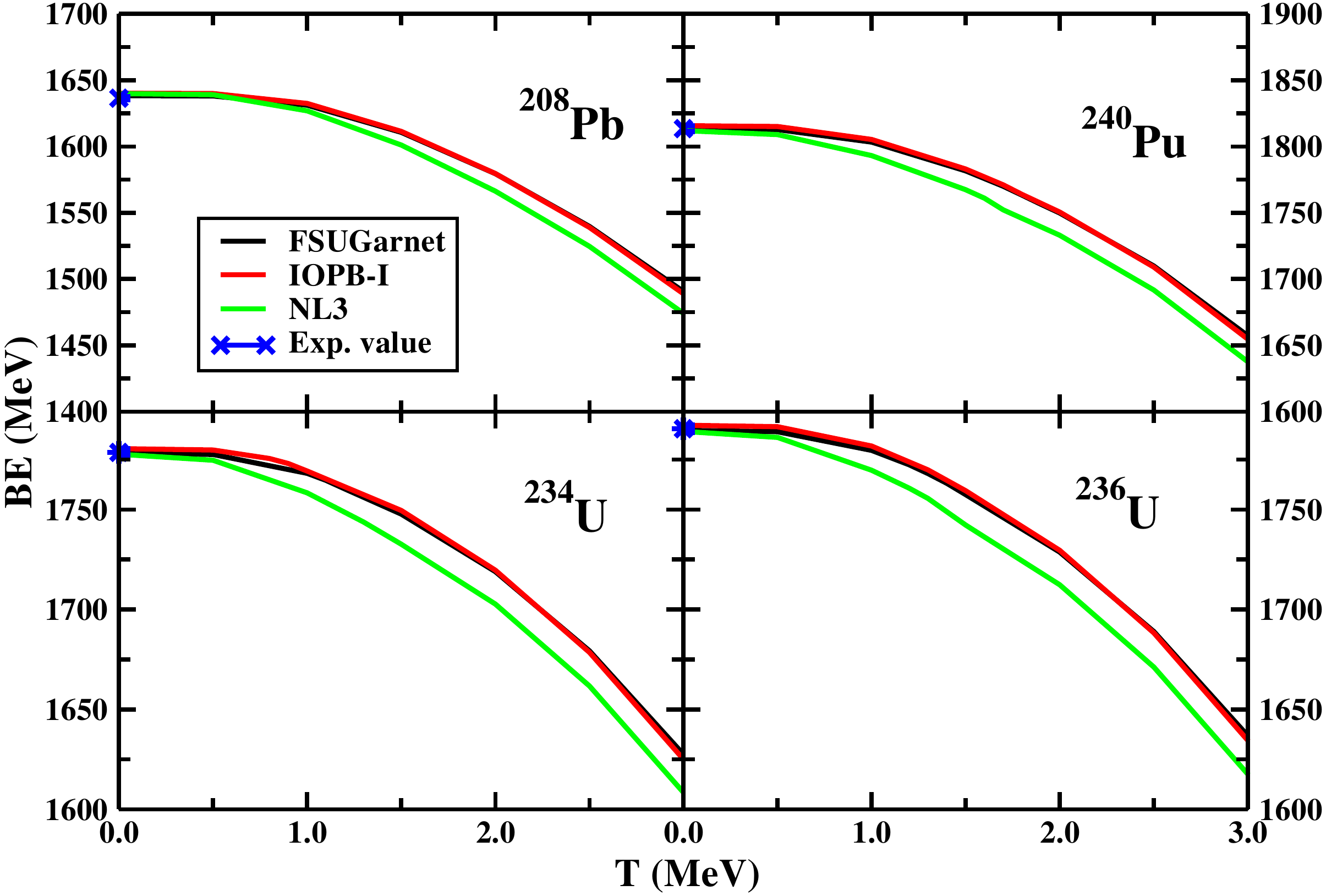}
        \label{Fig.1}\vspace{-0.6cm} \hspace{0.8cm} \caption{(Color online) The total binding 
energy (BE) as function of
        temperature T for $^{208}$Pb, $^{234}$U, $^{236}$U and $^{240}$Pu with FSUGarnet, IOPB-I and NL3 parameter
        sets.}
\end{figure}
\begin{figure}
	\hspace{0.8cm}
	\includegraphics[width=0.9\columnwidth,clip=true]{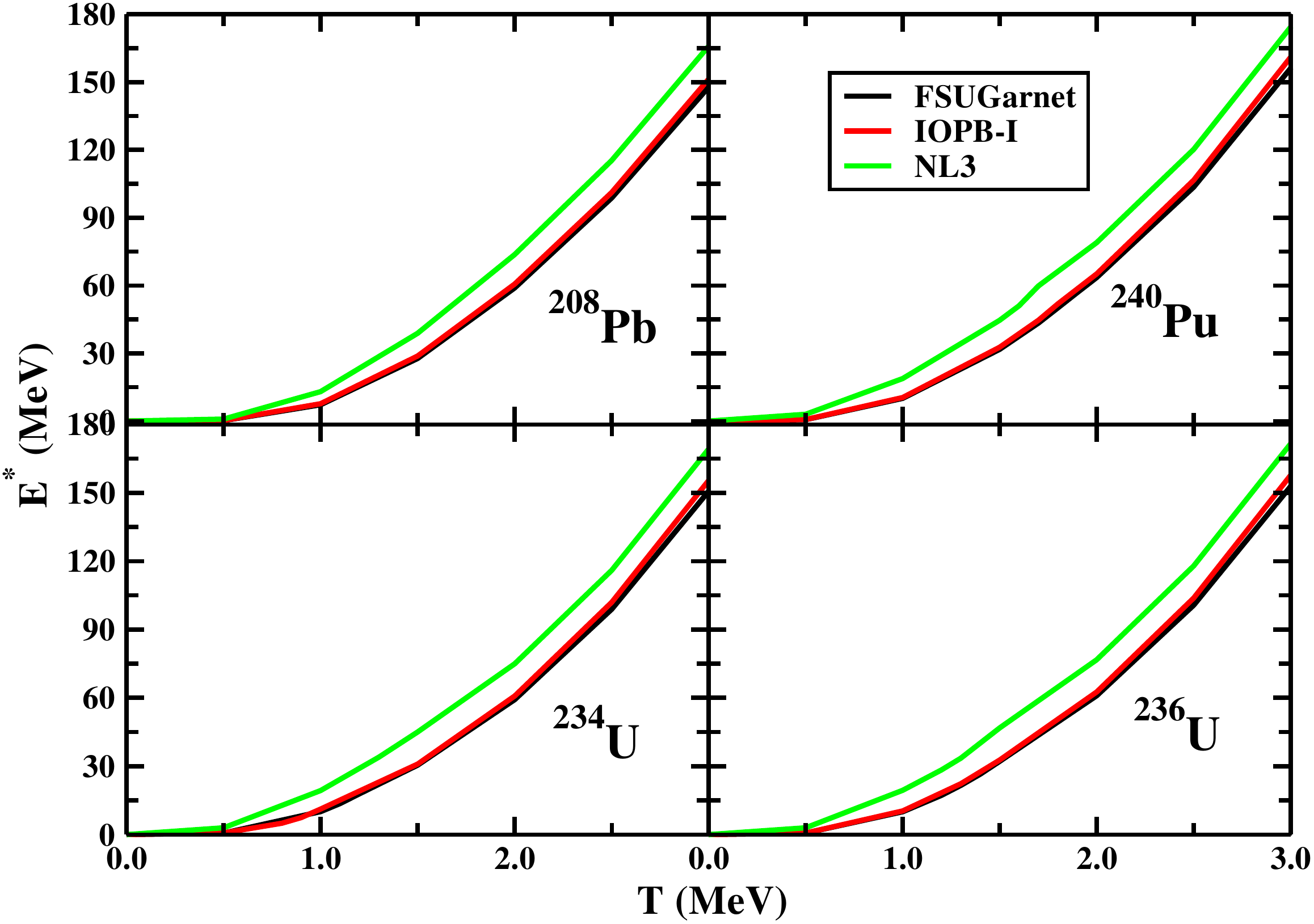}
	\label{Fig.1}\vspace{-0.6cm}\hspace{0.8cm} \caption{(Color online) The excitation energy $E^*$ as a function of 
        temperature T for $^{208}$Pb, $^{234}$U, $^{236}$U and $^{240}$Pu with FSUGarnet, IOPB-I and NL3 parameter
        sets.}
\end{figure}
\begin{figure}
	\hspace{0.8cm}
	\includegraphics[width=0.9\columnwidth,clip=true]{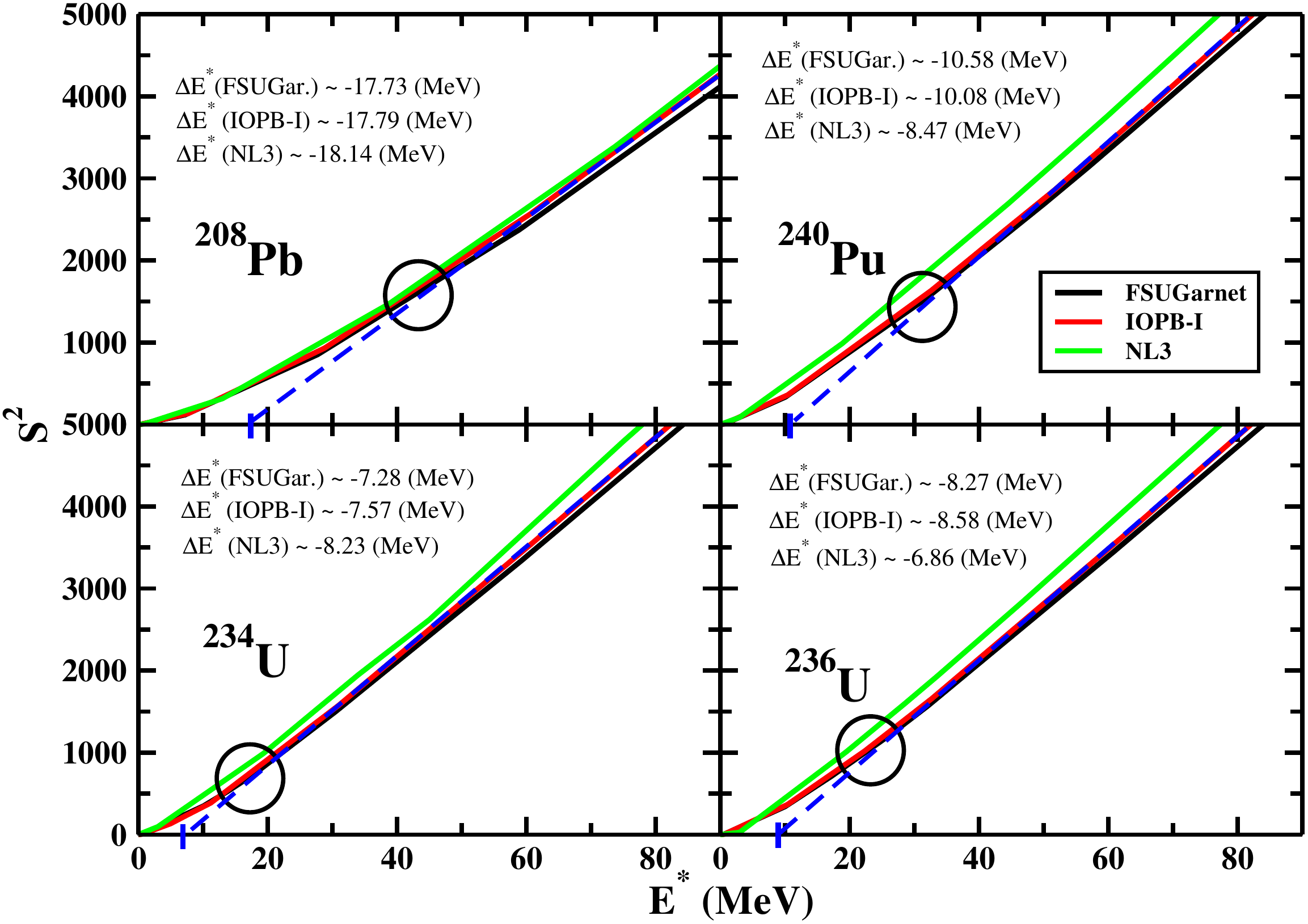}
	\label{Fig.9}\vspace{-0.6cm}\hspace{0.8cm} \caption{(Color online) The square of the entropy $S^2$ versus excitation 
        energy $E^*$ for $^{208}$Pb, $^{234}$U, $^{236}$U and $^{240}$Pu with FSUGarnet, IOPB-I and NL3 parameter 
        sets. The circle marked in the curve shows the shell melting point.}
\end{figure}

\subsection{Nuclear excitation energy and shell melting point}

After verifying the validity of these force parameters by studying the nuclear 
properties at ground state (T = 0 MeV), we have extended the calculations to study further at finite 
temperature. As the temperature of a nucleus rises, the nucleons are excited to
higher orbitals and the nucleus as a whole in an excited state. 
And hence, all its observables change with T.
Before going to study other properties of a nucleus, first we discuss the 
variation of its binding energy with T. The binding energies for $^{208}$Pb, 
$^{234,236}$U and $^{240}$Pu are shown in Figure 1 with FSUGarnet, IOPB-I and 
NL3 sets.
The binding energies in all the cases decrease gradually with the effect of
 temperature. It is found that at $T=0$, all the three forces give almost same 
binding energy. This is clear from both Table 2 and Figure 1 for 
the cases, where the experimental data are measured precisely ($T=0$). 
Also, the predicted 
results at $T=0$ coincide very much with the experimental data (see Figure 1 
and Table 2). Further, with the increase of temperature, the BE corresponding to 
NL3 force underestimates the prediction of other two sets IOPB-I and FSUGarnet
 as shown in the figure.

The nuclear excitation energy $E^*$ is one of the key quantity in fission dynamics. The excitation energy
very much depends on the state of the nucleus. It is defined as the nucleus excited how far from the ground
state and can be measured from the relation $E^*=E(T)-E(T=0)$, where E(T) is the binding energy of the
nucleus at finite T and E(T=0) is the ground state binding energy. The variation of $E^*$ as a function of T is 
shown in Figure 2 for $^{208}$Pb, $^{234,236}$U and $^{240}$Pu with NL3, FSUGarnet and IOPB-I sets.  
One can see from the figure that the variation of excitation energy is almost quadratic in nature satisfying the relation $E^* = aT^2$. 
similar to the binding energy, the results of $E^*$ for FSUGarnet and IOPB-I 
coincide with each other, but the $E^*$ predicted by NL3 set overestimate
these two as shown in Figure 2. 

The excitation energy $E^*$ has a direct relation
with the entropy S, i.e., the disorderness of the system. The expected relation of S with $E^*$ and the level 
density parameter $a$ from Fermi gas model is $S^2=4aE^*$ \cite{rama70}. 
However, this straight line relation of $S^2$ versus $E^*$
deviates at low excitation energy due to the shell structure of nucleus \cite{rama70}. The value of $S^2$
as a function of $E^*$ is shown in Figure 3 for the four considered nuclei. The intercept of the curve 
on the $E^*-$axis is a measure of shell correction energy to the nucleus. Thus the actual relation of $E^*$ with $S^2$ 
can be written as $S^2=4a(E^*\pm\triangle{E^*})$, where $\pm\triangle{E^*}=$ shell correction energy. Beyond these "slope
points" the $S^2$ versus $E^*$ curve increase in a straight line as shown in the figure (the slope point is marked by circle). 
Thus, one can interpret that beyond this particular
excitation energy, the nucleus as a whole does not have a shell structure, and can be considered as the melting
point of shell in the nucleus.  The value of this point depends on the ground state shell structure of the
nucleus. The, experimental evidences of washing out of the shell effects at and around 40 MeV excitation energy has also been pointed out in Ref 
\cite{Chaudhuri15}. Shell correction obtained from the intercepts on the $ E^*-$ axis in Figure 3 are depicted 
in the figure for all four nuclei 
corresponding to parameters set considered here. For example, shell correction energies are $\sim -17.79$, $-7.57$, $-8.58$ and $-10.08$ MeV 
for $^{208}$Pb, $^{234,236}$U and $^{240}$Pu respectively, corresponding to IOPB-I parameter set. 
The values for all three parameter sets are almost 
same with a little difference in NL3 model (see Figure 3).  

\begin{figure}
	\hspace{0.8cm}
	\includegraphics[width=0.9\columnwidth,clip=true]{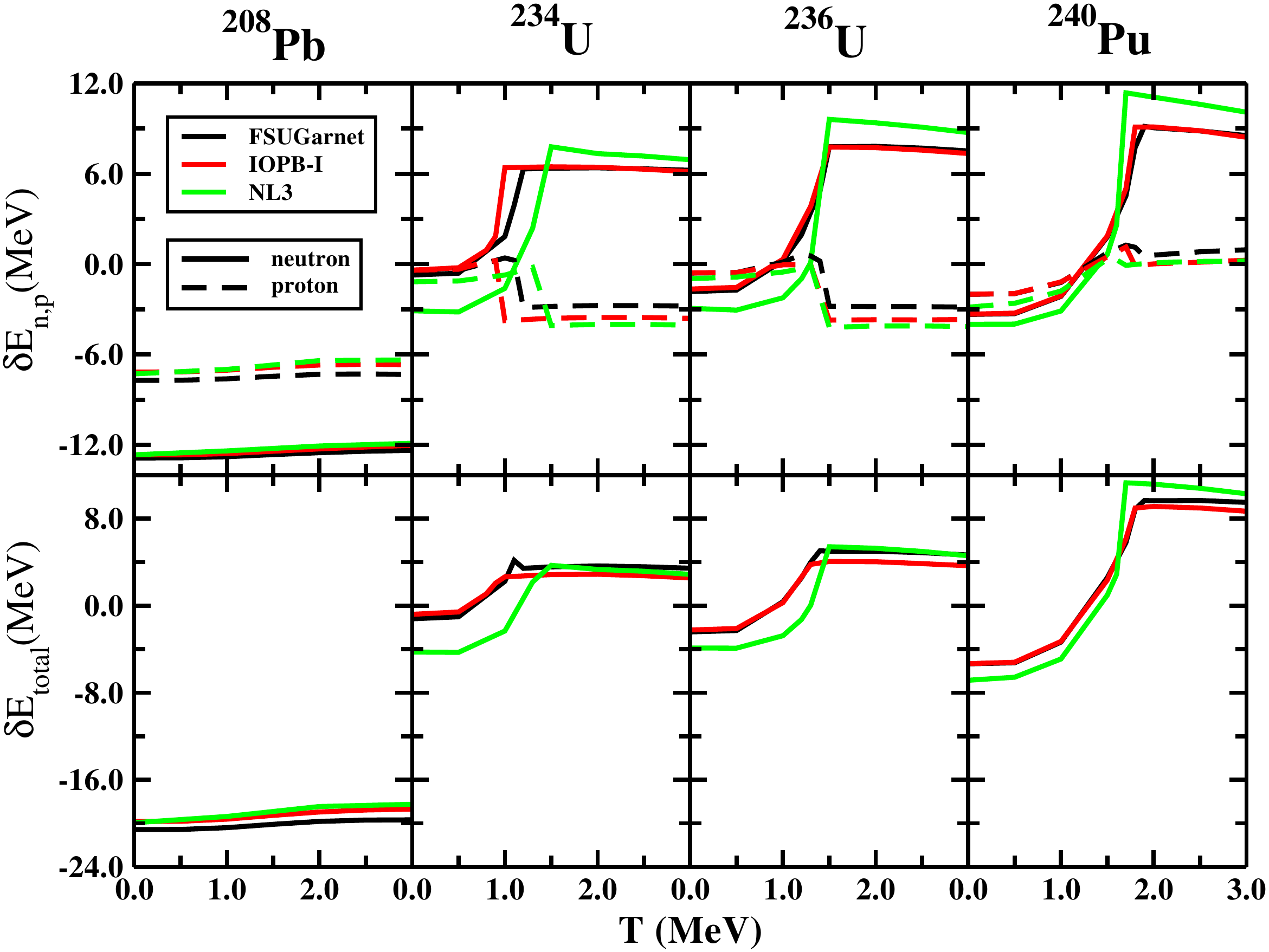}
	\label{Fig.4}\vspace{-0.6cm} \hspace{0.8cm} \caption{(Color online) The nuclear shell correction $\delta{E_{shell}}$ 
as a function of temperature T for the nuclei $^{208}$Pb,
        $^{234}$U, $^{236}$U and $^{240}$Pu with FSUGarnet, IOPB-I and NL3
        parameter sets.}
\end{figure}

A further analysis of the melting of shell, we calculate the nuclear shell correction energy $\delta{E_{shell}}$ for
protons, neutrons and total nucleons of the nucleus. This, we have evaluated in the frame-work of the well known 
Strutinsky shell correction prescription \cite{stru}. The relativistic single particle energies for protons
and neutrons at finite T obtained by various sets (NL3, FUSGarnet and IOPB-I) are the inputs in the calculations and
the results are depicted in Figure 4. As expected, the shell corrections (both protons and neutrons) for 
$^{208}$Pb, almost remains constant with temperature. Contrary to the behavior of $^{208}$Pb, the 
$\delta{E_{shell}}$ for $^{234,236}$U and $^{240}$Pu, initially increase with T for neutron up to the transition
point and then suddenly decrease monotonously. On the other hand, we get an abrupt change of $\delta{E_{shell}}$
for proton at the transition point and remains a constant value as shown in Figure 4. This transition point, i.e., the
shell melting point coincides with the results obtained from the $S^2\sim E^*$ curve (Figure 3). 

\begin{figure}
	\hspace{0.8cm}
	\includegraphics[width=0.9\columnwidth,clip=true]{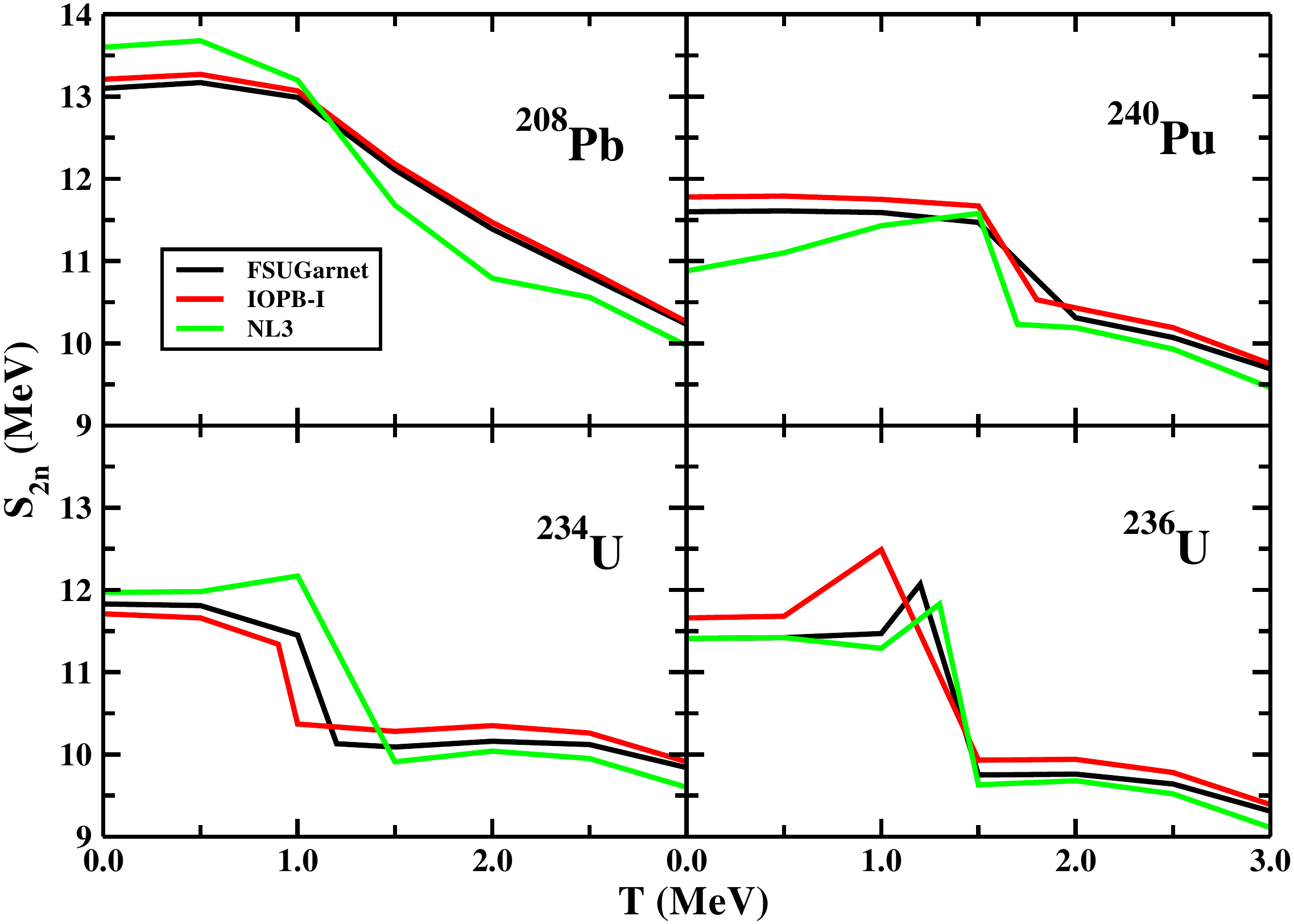}
	\label{Fig.7}\vspace{-0.6cm} \hspace{0.8cm}\caption{(Color online) The $S_{2n}-$separation energy as a function 
        of temperature T for the nuclei $^{208}$Pb, $^{234}$U, $^{236}$U and $^{240}$Pu with FSUGarnet, IOPB-I and NL3
        parameter sets. } 
\end{figure}

In Figure 5, we plot the two-neutron separation energy $S_{2n}(Z,N)=BE(Z,N)-BE(Z,N-2)$ with temperature T for $^{208}$Pb, 
$^{234}$U, $^{236}$U and $^{240}$Pu with FSUGarnet, IOPB-I and NL3 parameter sets. 
For $^{208}$Pb, there is no definite transition point in the 2n-separation energy.
In case of $^{236}$U, all forces give the same transition point (i.e., $T=1.5$ Mev), whereas in $^{234}$U and $^{240}$Pu
it is parameter dependent as shown in Figure 5.
 For example, in case of $^{234}$U we noticed the transition point
for NL3 a bit later than IOPB-I and FSUGarnet but for $^{240}$Pu, the transition point is opposite in nature (see Figure 5). 
In addition, for $^{208}$Pb, the curves corresponding to FSUGarnet and IOPB-I sets overlap and distinct from NL3. 
The $S_{2n}$ value decreases gradually for $^{208}Pb$. But, for other three nuclei this behavior is different. There is a sudden fall at the 
 transition temperature. Beyond these temperatures, the $S_{2n}$ value decreases
 almost smoothly. Thus, It can be concluded from the graph that the probability
 of emission of neutrons grows as temperature
 increases. This probability become very high at and beyond transition points.

\subsection{Single particle energy and shape transition}

\begin{figure}
	\hspace{0.8cm}
	\includegraphics[width=0.9\columnwidth,clip=true]{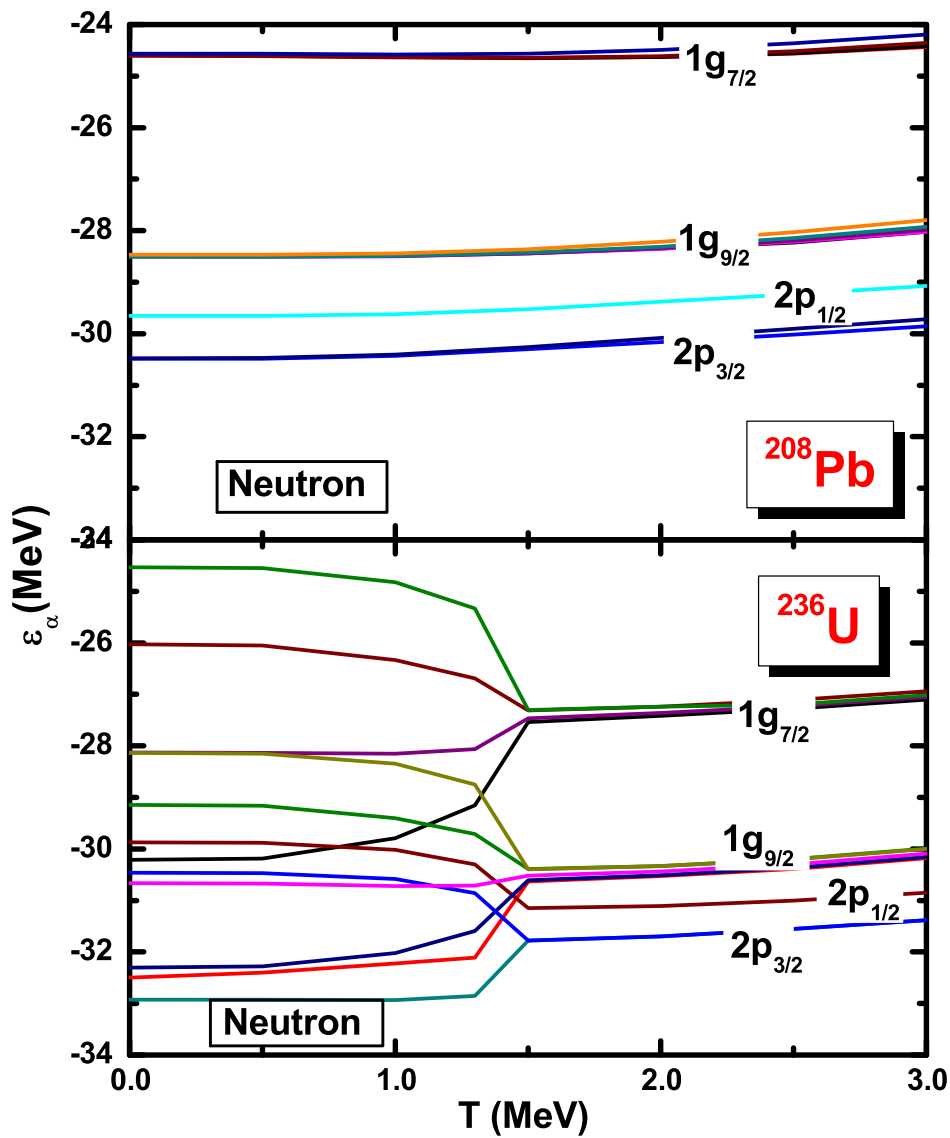}
	\label{Fig.9}\vspace{-0.6cm} \hspace{0.8cm}
        \caption{(Color online) The change of neutron single particle spectrum $\epsilon_{\alpha}$ for some selected levels
        as a function of temperature for the nuclei $^{208}$Pb, and $^{236}$U with IOPB-I parameter set.}
\end{figure} 

In Figure 6, we have shown a few single particle levels $\epsilon_{\alpha}$ of neutrons for the 
nuclei $^{208}$Pb and $^{236}$U as a representing case with IOPB-I set. From the figure, it is clear that the single particle energy 
of various $m$ states arising from different orbitals  merge to a single degenerate level with increase of temperature 
after a particular T for the $^{236}$U nucleus. We call this temperature $T=T_c$ as the shell melting or transition
temperature, because this is the temperature at which the nucleus changes from deformed to spherical state and all the
$m-$ levels converged to a single degenerate one \cite{Gambhir62,Agarwal64}. The same value of critical temperature is obtained 
in case of single particle spectrum of protons for the nuclei, i.e., non-degenerate levels become the degenerate. 
The same behavior is obtained for the other remaining thermally fissile nuclei considered here. 
When we compare this temperature with the shell melting
point (i.e. slope point of the $S^2$ vs $E^*$ curve), change in $\delta{E_{shell}}$ (Figure 4) and $S_{2n}$ values (Figure 5), the
merging point in $\epsilon_{n,p}$ matches perfectly with each other. We have 
increased the temperature further,
and analyzed the single particle levels at higher T, but we have not noticed the re-appearance of 
non-degenerate states. In case of $^{208}$Pb, the single particle energy for proton and neutron do not
change with temperature. This is because, there is no change in the shape of this nucleus as it is 
spherical throughout all the T. Although, the shell correction energy changes from negative 
to positive or vice versa, we 
have not found the disappearance of shell correction completely. This means, whatever be the temperature of the
nucleus, the shell nature remains there, of course the nucleons are in the degenerate state. So, disappearance of shell effects
implies the redistribution of shells at transition point i.e., from non-degenerate to degenerate. 
In other words, whatever be the temperature of the nucleus, there will be a 
finite value of shell correction energy (as shown in Figure 4) due to the random motion of nucleons. 
On inspecting the single-particle energy
for the entire spectrum (not shown in the figure), one can find that 
low lying states raise slightly and high lying states decrease slightly. This is due to an increase in the effective mass and 
rms radius \cite{Gambhir62}. 
We have repeated the calculations for other two parameter sets (NL3 and FSUGarnet) and find the same scenario.

\subsection{Quadrupole and Hexadecapole deformations and shape transition}

\begin{figure}
	\hspace{0.8cm}
	\includegraphics[width=0.9\columnwidth,clip=true]{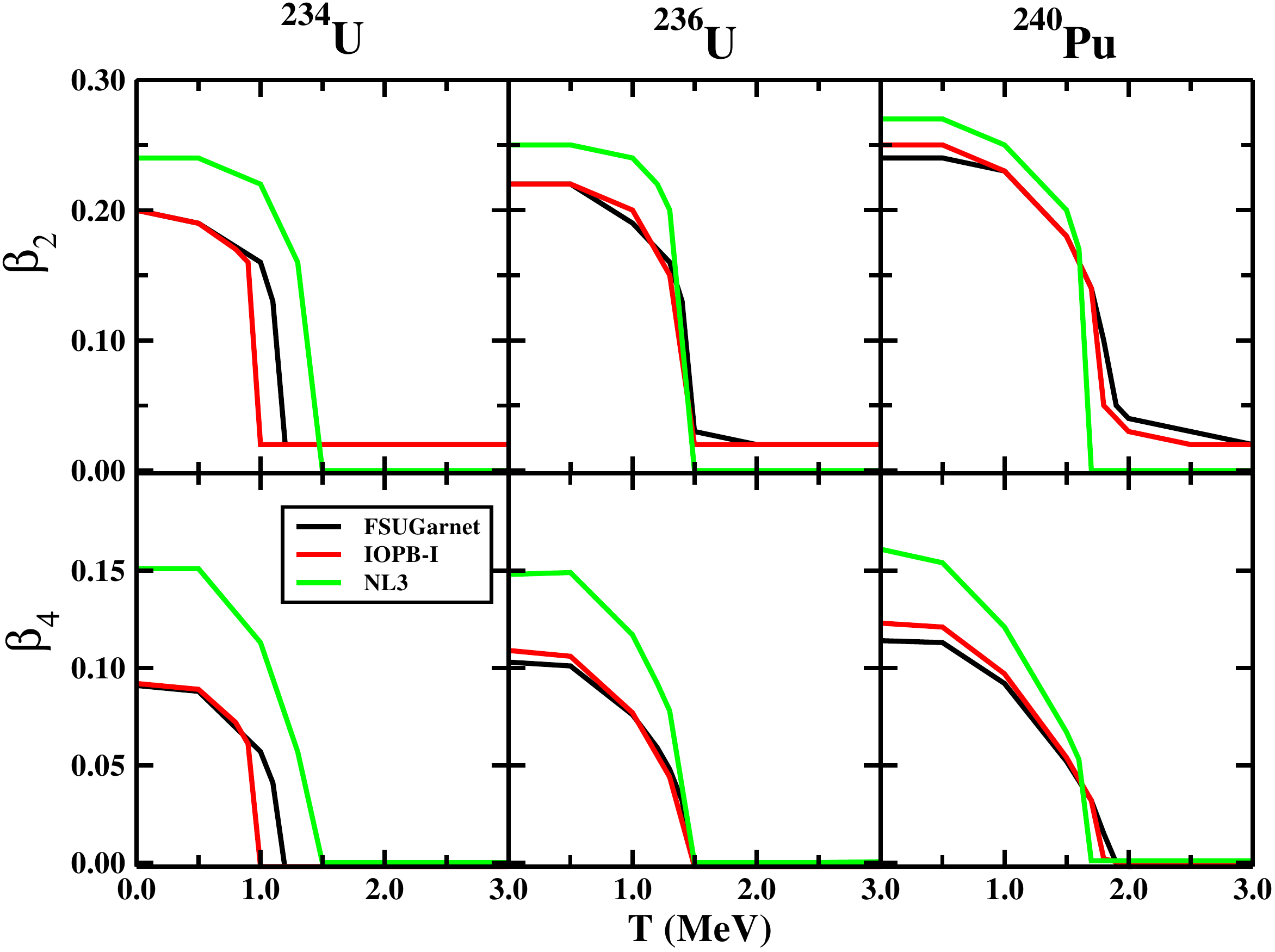}
	\label{Fig.2}\vspace{-0.6cm} \hspace{0.8cm}\caption{(Color online) The quadrupole and hexadecapole deformations parameters
        ($\beta_2$ and $\beta_4$,respectively) as a function of temperature T for $^{234}$U, $^{236}$U and $^{240}$Pu with 
        FSUGarnet, IOPB-I and NL3 parameter sets.}
\end{figure}

The quadrupole $\beta_2$ and hexadecapole $\beta_4$ deformation parameters for 
the nuclei 
$^{234}$U, $^{236}$U and $^{240}$Pu are shown in Figure 7. The upper panel is the $\beta_2$
and the lower panel is $\beta_4$ as a function of temperature T. Both the deformation parameters drastically
decrease with T at $T=T_c$. These results are qualitatively consistent with the previous studies 
for different nuclei \cite{Gambhir62,Agarwal62}. 
The almost zero value of $\beta_2$ at and beyond the critical temperature $T_c$ implies 
that the shape of nucleus changes form deformed to spherical. The non-smoothness on the surface 
of nucleus irrespective of its shape is defined by the  hexadecapole deformation parameter $\beta_4$. 
The same behavior for $\beta_4$ implies that 
on increasing temperature not only the shape of nucleus changes but also its surface becomes smooth.
Hence nucleus becomes a perfectly degenerate sphere at temperature $T_c$ and beyond. We have checked that even on further 
increasing temperature the state of nucleus does not change again and remain 
spherical as mentioned earlier. While, in fission process, a nucleus undergoes 
scission point where it is highly deformed and hence, breaks into fragments.
This observation concludes that the nucleus never undergoes fission only by temperature. 
It is necessary to disturb the nucleus physically for fission reaction bombarding thermal neutron.
In other words, $^{233}$U, $^{235}$U and $^{239}$Pu have half-lives $T_{1/2}=$
$1.59 \times 10^5$ years, $7.04 \times10^8$ years and $2.4 \times10^4$ years, respectively. This means,
 these nuclei never undergo fission spontaneously whatever be the temperature. 
However, the fission reaction takes place whenever a zero energy (thermal) neutron hits on it externally.

\subsection{Specific heat}

\begin{figure}
        \hspace{0.8cm}
        \includegraphics[width=0.9\columnwidth,clip=true]{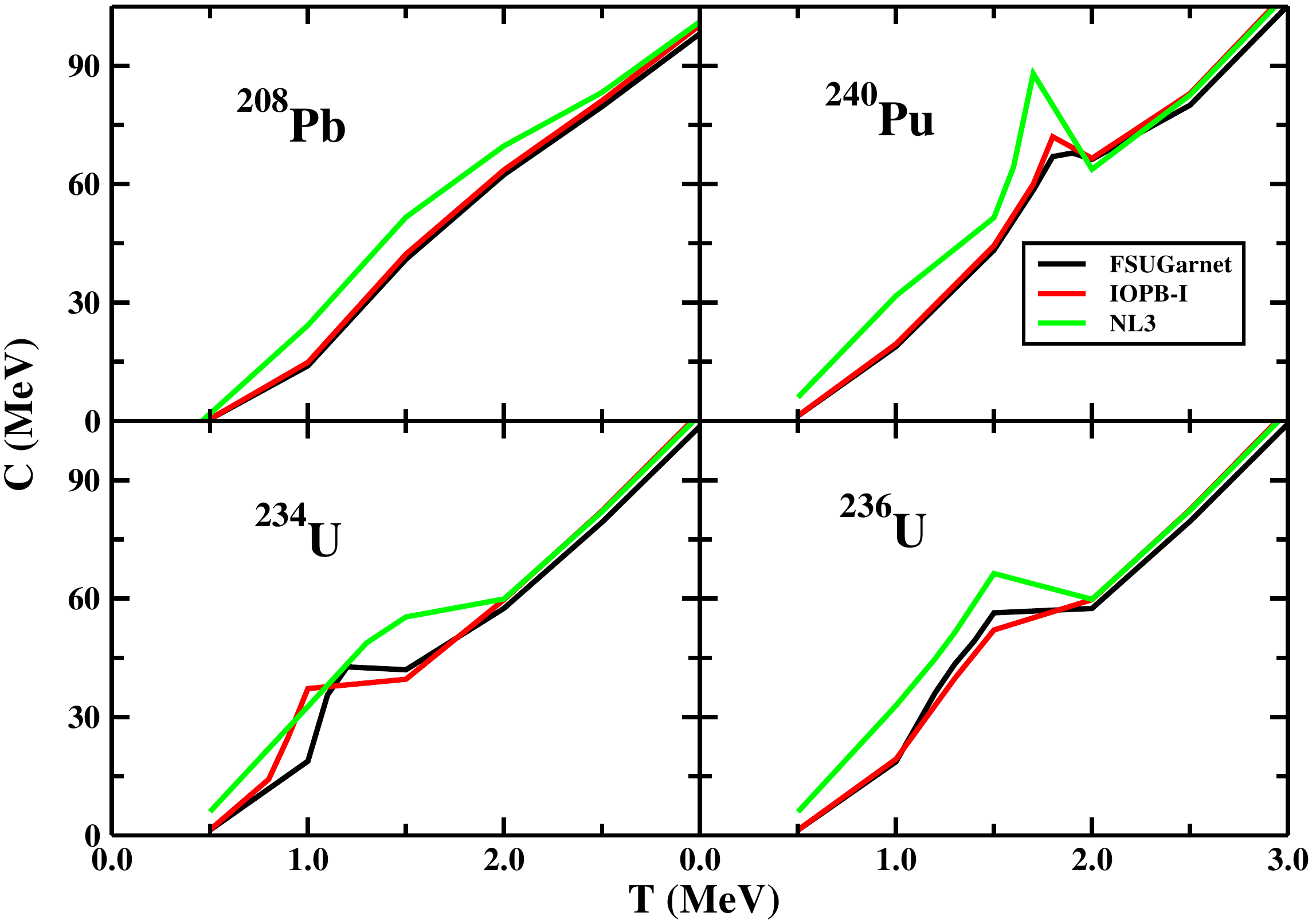}
        \label{Fig.3}\vspace{-0.6cm} \hspace{0.8cm}\caption{(Color online) The specific heat for the nuclei $^{208}$Pb,
        $^{234}$U, $^{236}$U and $^{240}$Pu as a function of temperature T for FSUGarnet, IOPB-I and NL3
        parameter sets.}
\end{figure}

For further study the phase transition at critical temperatures we have 
calculated specific heat for the nuclei considered here. The specific heat of a nucleus 
 is:

\begin{eqnarray}
C(T) & = & \frac{\partial E^*}{\partial T}\;.
\end{eqnarray}
The variation of specific heat with temperature is shown in Figure 8. The kinks in the curves 
for different nuclei correspond to the critical temperatures where shape transition takes place. 
This value differs with nucleus and in agreement with the obtained results shown in the previous graphs (Figures 3-7). 
But, it can be seen that for $^{208}Pb$, there is no such kink which signifies that 
it remains spherical at all temperatures. The pattern of the curves is same as studied earlier 
for $^{166}Er$ and $^{170}Er$ \cite{Agarwal62}. 
At low temperature FSUGarnet and IOPB-I results match and underestimates those of NL3 but at higher temperature i.e., 
beyond critical temperatures all the curves overlap with each others. 
The values of transition temperature are different for different parameter sets.

\subsection{Root mean square radius}

\begin{figure}
        \hspace{0.8cm}
        \includegraphics[width=0.9\columnwidth,clip=true]{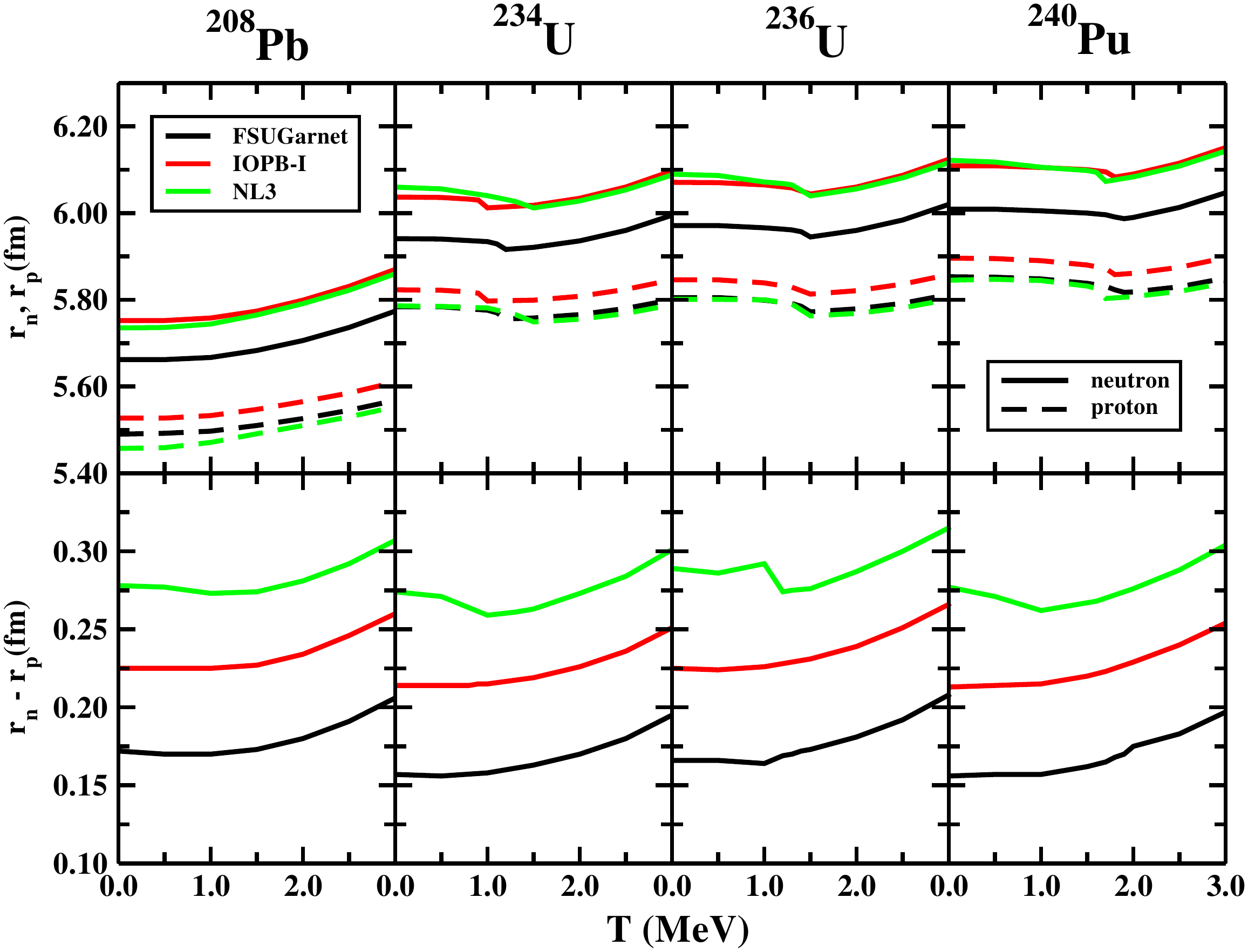}
        \label{Fig.6}\vspace{-0.6cm} \hspace{0.8cm}\caption{(Color online) The variation of root mean square (rms) neutron
        $r_n$ and proton $r_p$ radii as a function of temperature T for the nuclei $^{208}$Pb,
        $^{234}$U, $^{236}$U and $^{240}$Pu with FSUGarnet, IOPB-I and NL3
        parameter sets. (a) the upper panel is for $r_n$ and $r_p$ radii and (b) the lower panel is for neutron
        skin thickness $\triangle{r_{np}}=r_n-r_p$.}
\end{figure}

The root mean square radii for $^{208}$Pb, $^{234}$U, $^{236}$U and $^{240}$Pu are shown in Figure 9. 
The proton and neutron radii are presented in the upper panel whereas lower panel 
has the skin thickness for the nuclei. The rms radius, first increases slowly at lower temperature, and at higher 
temperature, it increases rapidly for $^{208}Pb$. This behavior is consistent with that discussed in Refs. \cite{Gambhir62,ant17}. 
For the remaining nuclei, the rms radius first decreases with temperature up to a point and beyond that point 
these values increase rapidly with temperature. This is because the deformed nucleus first undergoes 
phase transition to spherical shape and hence, radius decreases slightly 
and beyond the said point these increase rapidly. 
The numerical values of the transition temperatures are consistent with the shown values in the previous figures. 
It is also observed from the figure 
that neutron radius corresponding to IOPB-I and NL3 parameters matches with each other but overestimate that of FSUGarnet result. 
 For proton radius, FSUGarnet and NL3 results are coincide and underestimate that of IOPB-I. The values of rms radius are minimum at 
transition temperatures.
Behavior of skin thickness is also almost same as of nucleus radius. The kinks here correspond to the same transition temperatures.

There is a point to be discussed here that the neutron-skin thickness is an important quantity in the determination of EOS.
Although, there is a large uncertainty in the determination of $r_n$, even then some of the precise measurement are done 
 \cite{expskin}. Recently, it is reported \cite{fatto} using the Gravitational wave observation data GW170817 
that the upper limit of $\Delta r_{np}$ should be $\leq$ 0.25 fm for $^{208}Pb$. The calculated values of $\Delta r_{np}$ for NL3, 
FSUGarnet and IOPB-I are 0.28, 0.16 and 0.22 fm, respectively, which can be seen from Figure 9.  
The value of neutron-skin
thickness obtained by IOPB-I is preferred over NL3 and FSUGarnet. As we know, NL3 predicts a larger $\Delta r_{np}$
value which gives a larger neutron star radius \cite{iopb1}. Similarly, FSUGarnet predicts a smaller $\Delta r_{np}$ and expected 
a smaller neutron star radius. 

\subsection{Inverse level density parameter}

\begin{figure}
        \hspace{0.8cm}
        \includegraphics[width=0.9\columnwidth,clip=true]{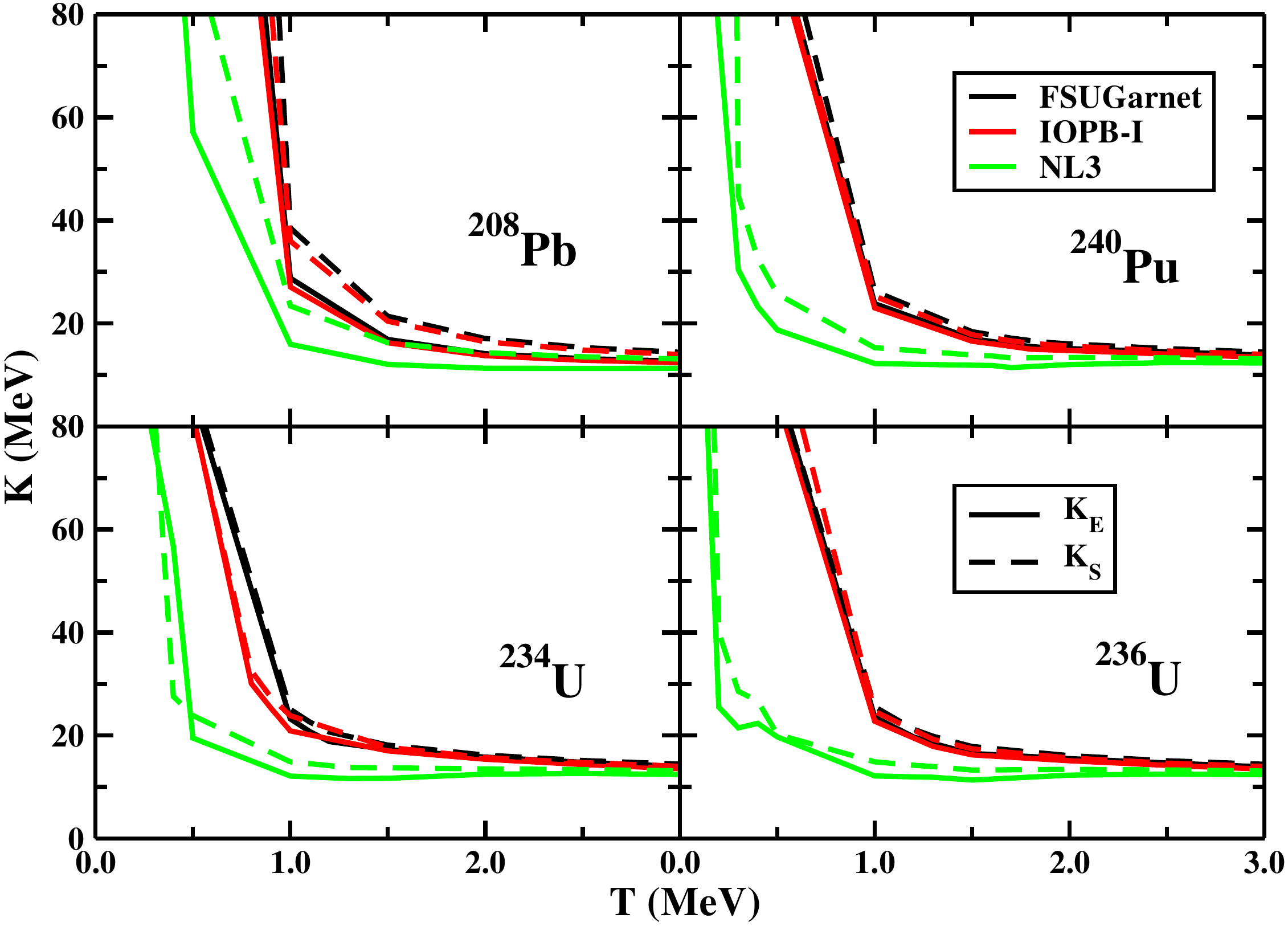}
        \label{Fig.5}\vspace{-0.6cm} \hspace{0.8cm}\caption{(Color online) The inverse level density parameter
        as a function of temperature T for the nuclei $^{208}$Pb,
        $^{234}$U, $^{236}$U and $^{240}$Pu with FSUGarnet, IOPB-I and NL3
        parameter sets.}
\end{figure}

Level density parameter is a key quantity in the study of nuclear fission \cite{bethe}. 
It is used to predict the probability of fission fragments, 
yield of fragments and ultimately angular distribution of fragments \cite{seinthil17,bharat17}. The motivation 
behind choosing thermally fissile nuclei as discussed above is to study fission parameters at finite temperature which may be 
useful for theoretical and experimental fission study. 
Dependence of level density parameter on temperature and its sensitivity to parameters chosen are discussed below. 
It can be obtained using the excitation energy and entropy as follows:

\begin{eqnarray}
E^* & = & aT^2 \:,
\label{eq34}
\end{eqnarray}
\begin{eqnarray}
S & = & 2aT \:.
\label{eq35}
\end{eqnarray}

The parameter $a$ obtained from equations (\ref{eq34}) and (\ref{eq35}) are equal, when it is independent of 
temperature \cite{Agarwal98}. This is true for higher temperature i.e., beyond critical point. But, at low 
temperature, it is quite sensitive to T. The inverse level density parameter 
defined as $ K = A/a$ (where A is the mass of the nucleus) is presented in Figure 10. The bold and dashed lines 
correspond to the K values 
obtained by using the above two equations (\ref{eq34}) and (\ref{eq35}) are represented by the symbols $ K_E $ and $ K_S$. 
At low temperature, values of K shoot up and then it becomes smooth with the variation of temperature. The pattern of $ K_E$ and $K_S $ 
are same and there is no appreciable change as temperature rises except the kinks which again correspond to critical points. 
The constant K shows the vanishing of shell structure of the nucleus at higher temperature (See Fig. 10). 
These excitation energies are consistent with those shown in Figure 3 
wherein slope of $ S^2$ versus $ E^*$ curve representing level 
density parameter. It is clear from the figure that FSUGarnet and IOPB-I curves overlap with each other and overestimate 
the values obtained by NL3 model. This sensitivity of level density parameter on the choice of Lagrangian density 
will affect the fragment distributions in the fission process.

\subsection{Asymmetry energy coefficient}

\begin{figure}
        \hspace{0.8cm}
        \includegraphics[width=0.9\columnwidth,clip=true]{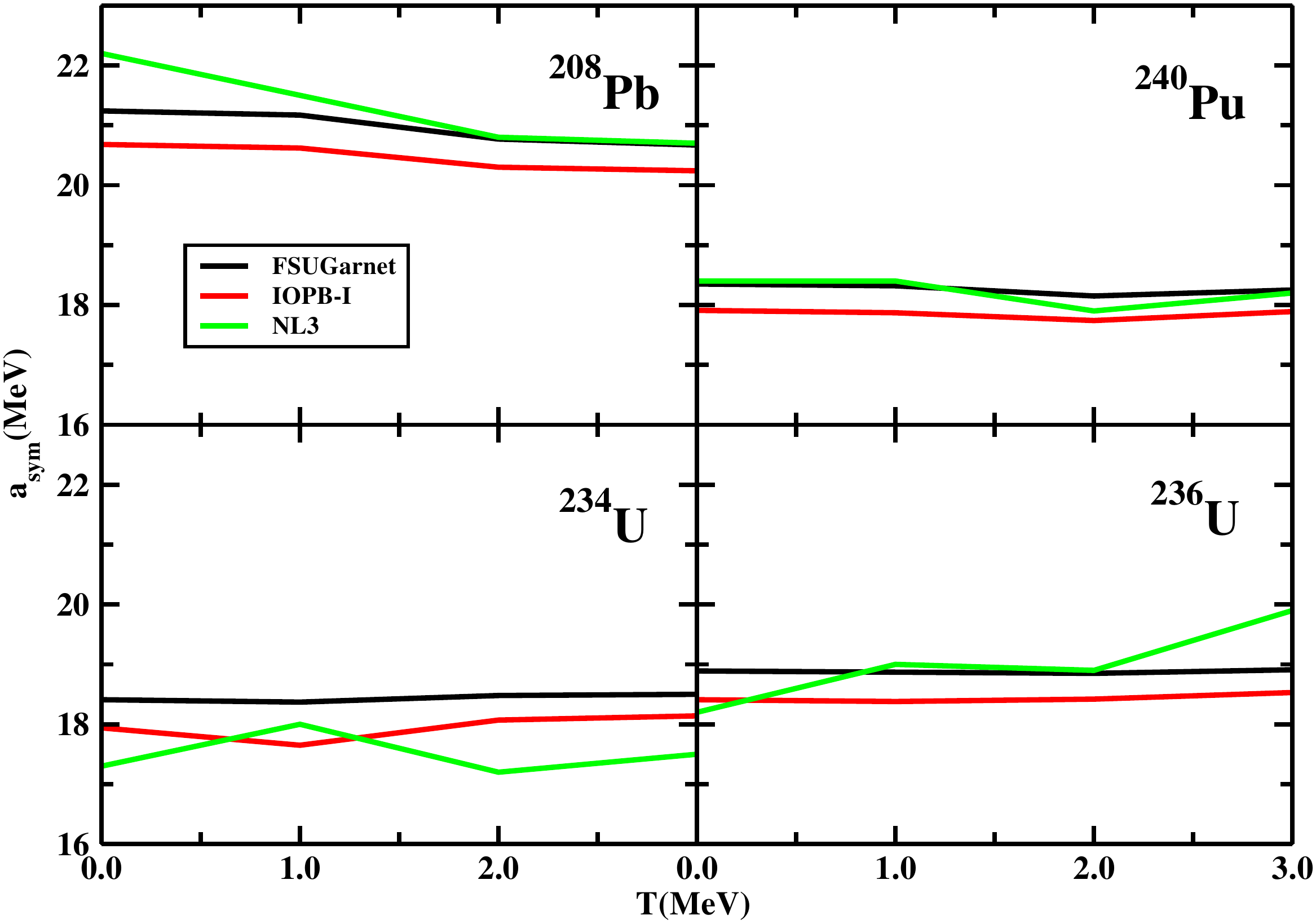}
        \label{Fig.8}\vspace{-0.6cm} \hspace{0.8cm}\caption{(Color online) The asymmetry energy coefficient $a_{sym}$ as a function
        of temperature T for the nuclei $^{208}$Pb, $^{234}$U, $^{236}$U and $^{240}$Pu with FSUGarnet, IOPB-I and NL3
        parameter sets.}
\end{figure}

The properties of hot nuclei play a vital role in both nuclear physics and astrophysics \cite{shlomo05}. These properties are 
the excitation energy, entropy, symmetry energy, and density distribution $etc.$. Among these properties, symmetry energy 
and its dependence on density and temperature have a crucial role in understanding various phenomena in heavy ion collision, 
supernovae explosions, and neutron star \cite{barran,prakash}. It is a measure of energy gain in converting isospin 
symmetric nuclear matter to asymmetric one. The temperature dependence of symmetry energy plays its role in changing the location 
of neutron drip line. It has key importance for the liquid-gas phase transition of asymmetric nuclear matter, 
the dynamical evolution of mechanism of massive star and the supernovae explosion \cite{barron85}.

Experimentally, nuclear symmetry energy is not a directly measurable quantity. It is extracted indirectly from the observables 
those are related to it \cite{Shetty,kowalski}. Theoretically, there are different definition for different systems. 
For infinite nuclear matter, symmetry energy is the expansion of  binding energy per nucleon in terms of isospin asymmetry 
parameter, i.e., $ \alpha = (\rho_n - \rho_p)/\rho $ as \cite{DeA12}:

\begin{eqnarray}
\epsilon (\rho,\alpha) & = & \epsilon (\rho,0) + a^{\upsilon}_{sym} (\rho)\alpha^2 + O(\alpha^4) \:, 
\end{eqnarray}
where $\rho = \rho_n + \rho_p $ is the nucleon number density. The coefficient $a^{\upsilon}_{sym}$ in the second term is the asymmetry 
energy coefficient of nuclear matter at density $\rho$.   
The value of $a_{sym}$ for nuclear matter at saturation density $\rho_0$ is about 30-34 MeV \cite{DeA12,Agarwal14,hard06}. 
For finite nuclei, symmetry energy is defined as one of the contributing term due to asymmetry in the system 
in the Bethe-Weizsacker mass formula. The coefficient of symmetry energy is defined as \cite{Daniel03}

\begin{eqnarray}
a_{sym}(A) & = & \frac{a^{\upsilon}_{sym}}{1 + (a^{\upsilon}_{sym}/a^s_{sym})A^{-1/3}}
\;,
\end{eqnarray}
where $ a^{\upsilon}_{sym}$ and $ a^s_{sym}$ are the volume and surface asymmetry energy coefficients. The 
$a^{\upsilon}_{sym}$ is considered as the asymmetry energy coefficient of infinite nuclear matter at saturation density.  
Here, we have calculated asymmetry energy coefficient of the nucleus of mass number A at finite temperature by using 
the method \cite{dean02,De12}:

\begin{eqnarray}
a_{sym}(A,T) = [\epsilon_b(A,X_1,T) - \epsilon_b(A,X_2,T)]/(X^2_1-X^2_2)
\;,
\nonumber \\
\end{eqnarray}
where $X_1$ and $X_2$ are the neutron excess ($X=\frac{N-Z}{A}$) of a pair 
of nuclei having same mass number A but different proton number Z. 
We have taken $Z_2 = Z_1-2$ for calculating $a_{sym}$ where, 
$Z_1$ is the atomic number of the considered nucleus. 
The $\epsilon_b$ 
is the energy per particle obtained by subtracting Coulomb part, i.e., Coulomb 
energy due to exchange of photon in the interaction of nucleons is subtracted
from the total energy of the nucleus ($\epsilon_b = \epsilon - \epsilon_c$).  
For example, to study the asymmetry energy coefficient $a_{sym}$ for $^{236}$U at temperature T
using Eq. (34), we estimate the binding energy $\epsilon_b(A,X_1,T)$ for $^{236}$U
at finite temperature T without considering Coulomb contribution. Then the binding
energy of $^{236}$Th $\epsilon_b(A,X_2,T)$ is measured in the similar conditions, i.e.
without Coulomb energy at temperature T. Here $Z_1=92$ for Uranium and $Z_2=90$ for Thorium with
same mass number A=236 are chosen. Note that the value of $Z_2$ need not be $Z_1-2$, but
generally a pair of even-even nuclei are considered whose atomic number differer by 2.
The temperature dependence $a_{sym}$ coefficient  for different isotopic chains had 
been studied by using various definition as mentioned above and with relativistic 
and non-relativistic Extended Thomas Fermi Model \cite{ant17,De12,Zhang14}.
In Ref. \cite{De12}, the effect of choosing different pairs of nuclei on $a_{sym}$ coefficient is discussed.

The asymmetry energy coefficient $a_{sym}$ for all four nuclei are shown in Figure 11. From the figure, one can not say that the asymmetry 
energy co-efficient follows a particular pattern. This value is highly force dependent and results of different nuclei
have different trend with temperature. For example, in case of $^{208}$Pb, IOPB-I set consistently predicts smaller $a_{sym}$
than those of NL3 and FSUGarnet. At low temperature, NL3 has the larger $a_{sym}$ even than FSUGarnet, but for 
relatively high temperature $a_{sym}$ of both NL3 and FSUGarnet coincide with each others. In case of $^{234}$U, the scenario is 
completely different. Here, FSUGarnet consistently predicts a larger $a_{sym}$ than NL3 and IOPB-I. But, the values of NL3 
and IOPB-I crosses each other at various places with increase of T. In case of $^{240}$Pu, although $a_{sym}$ is smaller 
for IOPB-I, but magnitude-wise all the three results are almost similar to each other. All the three sets predict very different 
results with each other for $^{236}$U; FSUGarnet gives a nearly constant value throughout all the temperature, IOPB-I 
initially predicts constant behavior and suddenly decreases with T, but NL3 gives a consistently increasing value with T.

\section{Summary and conclusions}

In summary, for the first time we have applied the recently proposed FSUGarnet and IOPB-I parameter sets 
of E-RMF formalism to deform nuclei at finite temperature. The bulk properties of finite nuclei, such as binding energy, 
charge radius, quadrupole and hexadecapole deformation parameters are evaluated. In this context, we have 
chosen the $^{234,236}$U and $^{240}$Pu nuclei, because these compound nuclei are formed by
absorbing a thermal neutron from their thermally fissile parents $^{233,235}$U and $^{239}$Pu, respectively. Our 
calculated results are compared with 
the prediction of the well known NL3 parameter set as well as with experimental data, wherever available. These 
properties for hot nuclei are of great importance in both nuclear fission and astrophysics. 
A detail analysis is also performed on excitation energy, shell structure including shell correction and melting of
shell with temperature. The quadrupole and hexadecapole shape change with the effect of temperature  
 are discussed elaborately. We found that $\beta_2$ and $\beta_4$ decrease
with temperature and finally the nucleus attains a degenerate Fermi liquid. The nuclear asymmetry energy 
coefficient is a crucial quantity both for finite nucleus and
nuclear matter. Here, we have studied the asymmetry energy coefficient with temperature and found that the
coefficient has a diverse nature. It depends on the parameter sets and the nuclear systems.

\section{Acknowledgment}
The authors thank Bharat Kumar, P. Arumugam and B. K. Agrawal for fruitful discussions. 
AQ and KCN acknowledge Institute of Physics (IOP), Bhubaneswar for providing the necessary computer facility and
hospitality. Abdul Quddus thanks to Department of Science and Technology, Govt. of India for the partial support in the form of INSPIRE fellowship.
Partial financial support is also provided by the Department of Science and Technology, Govt. of India, Project No. EMR/2015/002517.


\section*{References}

\begin{thebibliography}{190}
\bibitem{nndc} National {\it Nucl. Data Center}, www. nndc. bnl. gov.
\bibitem{otto} O. Hahn and F. Strassmann, Naturwissenschaften {\bf26}, 755 (1938).
\bibitem{bharat17}
              Bharat Kumar, M. T. Senthil Kannan, M. Balasubramaniam, B. K. Agrawal, and S. K. Patra,
              Phys. Rev. C {\bf 96}, 034623 (2017).
\bibitem{seinthil17}
              M. T. Senthil Kannan, Bharat Kumar, M. Balasubramaniam, B. K. Agrawal, and S. K. Patra,
              Phys. Rev. C {\bf 95}, 064613 (2017).
\bibitem{brown} B. A. Brown, Phys. Rev. Lett. {\bf85}, 5296 (2000).
\bibitem{horo86} C. J. Horowitz and J. Piekarewicz, Phys. Rev. Lett. {\bf86}, 
5647 (2001).
\bibitem{barran}
              V. Baran, M. Colonna, V. Greco, and M. Di Toro, Phys. Rep.
              {\bf 410}, 335 (2005).
\bibitem{prakash}
              J. M. Lattimer and M. Prakash, Phys. Rep.
              {\bf 442}, 109 (2007).
\bibitem{barron85}
              E. Baron, J. Cooperstein, and S. Kahana, Phys. Rev. Lett.
              {\bf 55}, 126 (1985).
\bibitem{fong71} P. Fong, Phys. Rev. C {\bf 3}, 2025 (1971).
\bibitem{chenya88}
              V. A. Rubchenya and S. G. Yavshits, Z. Physik A: At. Nucl.
{\bf 329}, 217 (1988).
\bibitem{lest04} 
              J. P. Lestone, Phys. Rev. C {\bf 70}, 021601 (2004).
\bibitem{fong56} P. Fong, Phys. Rev. {\bf 102}, 434 (1956).
\bibitem{maran09} 
              K. Manimaran and M. Balasubramanian, Phys. Rev. C {\bf 79},
024610 (2009).
\bibitem{toke} J. Toke, et. al.,  Nucl. Phys. A {\bf 440}, 327 (1985).
\bibitem{bord} B. Borderie, et. al., Z Physik A: At. and Nucl. {\bf 299}, 
263 (1981).
\bibitem{rama85} V. S. Ramamurthy and S. S. Kapoor, 
                Phys. Rev. Lett. {\bf 54}, 178 (1985).
\bibitem{swiat} W J Swiatecki,  Physica Scr. {\bf 24}, 113 (1981).
\bibitem{diehl74} H. Diehl and W. Greiner, Nucl. Phys. A {\bf 229}, 29 (1974).
\bibitem{wilkins}
              B. D. Wilkins, E. P. Steinberg, and R. R. Chasman,
 Phys. Rev. C {\bf 14}, 1832 (1976).
\bibitem{pauli} H. C. Pauli and T. Ledergerver,  Nucl. Phys. A
{\bf 175}, 545 (1971).
\bibitem{herbert} Herbert Diehl and Walter Greiner, Nucl. Phys. A. 
{\bf 229}, 29 (1974). 


\bibitem{bender}
               M. Bender, P.-H. Heenen, and P.-G. Reinhard, Rev. Mod. Phys. {\bf 75}, 121 (2003).
\bibitem{bogu77}  
             J. Boguta and A. R. Bodmer, Nucl. Phys. A  {\bf 292}, 413 (1977).

\bibitem{vat70}
              D. Vautherin and D. M. Brink, Phys. Lett. B \textbf{32}, 149 (1970); Phys. Rev. C \textbf{5}, 
              626 (1972).
\bibitem{pal} 
              M. K. Pal and A. P. Stamp, Nucl. Phys. A \textbf{99},  228 (1967).
\bibitem{skp10} 
              S. K. Patra, R. K. Choudhury and L. Satpathy, J. Phys. G.  \textbf{37},  085103 (2010).
\bibitem{bharat95} Bharat Kumar, S. K. Biswal and S. K. Patra,
                   Phys. Rev. C {\bf95}, 015801 (2017)
\bibitem{chai15}
              Wei-Chai Chen and J. Piekarewicz, Phys. Lett. B {\bf 748}, 284 (2015).
\bibitem{iopb1}
              Bharat Kumar, B. K. Agrawal and S. K. Patra, arXiv:1711.04940v2
\bibitem{lala97}
              G. A. Lalazissis, J. K\"onig and P. Ring, Phys. Rev. C {\bf 55}, 540 (1997).

\bibitem{blaizot} J. P. Blaizot, Phys. Rep. {\bf 64}, 171 (1980).

\bibitem{furnstahl97} R. J. Furnstahl, B. D. Serot and H. B. Tang,
              Nucl. Phys. A {\bf 598}, 539 (1996); R. J. Furnstahl, B. D. Serot and H. B. Tang,
              Nucl. Phys. A {\bf 615}, 441 (1997).
\bibitem{bharatnpa} Bharat Kumar, S. K. Singh, B. K. Agrawal, and S. K. Patra,
                   Nucl. Phys. A {\bf 966}, 197 (2017).
\bibitem{muller96} 
              H. M\"uller and B. D. Serot, Nucl. Phys. A {\bf 606}, 508  (1996).
\bibitem{serot97}  
              B. D. Serot and J. D. Walecka, Int. J. Mod. Phys. E {\bf 6}, 515 (1997).
\bibitem{estal01} 
              M. Del Estal, M. Centelles, X. Vi\~nas and S. K. Patra, Phys. Rev. C {\bf 63}, 024314 (2001).
\bibitem{negele} J. W. Negele, Phys. Rev. C {\bf 1}, 1260 (1970). 
\bibitem{blunden87} 
              P. G. Blunden  and  M. J. Iqbal, Phys.  Lett.  B  {\bf 196}, 295 (1987).
\bibitem{reinhard89} 
              P. G. Reinhard,  Rep.  Prog.  Phys. {\bf 52}, 439 (1989).
\bibitem{gam90} 
              Y. K. Gambhir, P. Ring and A. Thimet, Ann. of Phys. {\bf 198}, 132 (1990).

\bibitem{bohr} A. Bohr, B. R. Mottelson and D. Pines, 
               Phys. Rev. {\bf 110}, 4 (1958).

\bibitem{patra93}
              S. K. Patra, Phys. Rev. C  {\bf 48}, 1449 (1993).
\bibitem{pres82}
              M. A. Preston and R. K. Bhaduri, {\it Structure of Nucleus, Addison-Wesley Publishing Company},
              Ch. 8, page 309 (1982).
\bibitem{va73} 
              D. Vautherin, Phys. Rev. C \textbf{7}, 296 (1973).
\bibitem{dech80} 
              J. Decharg\'e and D. Gogny, Phys. Rev. C  {\bf 21}, 1568 (1980).
\bibitem{dutra12} M. Dutra, O. Lourenco, S. S. Avancini, B. V. Carlson, 
A. Delfino, D. P. Menezes, C. Providencia, S. Typel, and J. R. Stone, 
Phys. Rev. C {\bf 90}, 055203 (2014).
\bibitem{fujita} J. I. Fujita and H. Miyazawa, Prog. Theor. Phys. {\bf 17}, 360 (1957). 
\bibitem{pie} S. C. Pieper, V. R. Pandharipande, R. B. Wiringa, and J. Carlson,
                Phys. Rev. C {\bf 64}, 014001 (2001).
\bibitem{serot86} B. D. Serot and J. D. Walecka, Adv. Nucl. Phys.  {\bf 16}, 1 (1986).
\bibitem {wal74} 
              J. D. Walecka, Ann. Phys.  {\bf 83}, 491 (1974).
\bibitem{toki} Y. Sugahara and H. Toki, Nucl. Phys. A {\bf 579}, 557 (1994).
\bibitem{pika05}
              B. G. Todd-Rutel and J. Piekarewicz, Phys. Rev. Lett. {\bf 95}, (2005) 122501;
              C. J. Horowitz and J. Piekarewicz, Phys. Rev. Lett.  {\bf 86}, (2001) 5647;
              Phys. Rev. C {\bf 64} (2001), 062802 (R).
\bibitem{Angeli2013} I. Angeli, K. P. Morinova {\it Atomic Data and Nuclear Data Table}
              {\bf 99}, 69 (2013).
\bibitem{rama70}
              V. S. Ramamurthy, S. S. Kapoor and S. K. Kataria, Phys. Rev. Lett. {\bf 25}, 386 (1970).
\bibitem{Chaudhuri15} 
              A. Chaudhuri, et al., Phys. Rev. C {\bf 91}, 044620 (2015). 
\bibitem{stru} V. M. Strutinsky,  Nucl. Phys.
              {\bf A95}, 420 (1967).
\bibitem{Agarwal64} 
              B. K. Agrawal, Tapas Sil, S. K. Samaddar, and J. N. De, Phys. Rev. C
              {\bf 64}, 017304 (2001).
\bibitem{Gambhir62}
              Y. K. Gambhir, J. P. Maharana, G. A. Lalazissis, C. P. Panos and P. Ring, Phys. Rev. C
              {\bf 62}, 054610 (2000).
\bibitem{Agarwal62} 
              B. K. Agrawal, Tapas Sil, J. N. De, and S. K. Samaddar, Phys. Rev. C
              {\bf 62}, 044307 (2000).
\bibitem{ant17} 
              A. N. Antonov, D. N. Kadrev, and et al., Phys. Rev. C
              {\bf 95}, 024314 (2017).
\bibitem{expskin}
               A. Trzci\'nska,  J. Jastrzebski,  P. Lubi\'nski,  F. J. Hartmann,  R.Schmidt,  T.  von  Egidy,  and  B.  Klos,
               Phys.  Rev.  Lett. {\bf 87}, 082501 (2001);
               J. Jastrzebski,  A. Trzci\'nska,  P. Lubi\'nski,  B.Klos, F. J. Hartmann, T. von Egidy, and S. Wycech,
               Int. J. Mod. Phys. E {\bf 13}, 343 (2004).
\bibitem{fatto} F. J. Fattoyev, J. Piekarewicz and C. J. Horowitz, arXiv: 1711.06615v1; 
                P. B. Abbott et. al., Phys. Rev. Lett. {\bf 119}, 161101 (2017).
\bibitem{bethe} H. A. Bethe, Nuclear Phys. B. {\bf 9}, 69 (1937).
\bibitem{Agarwal98} 
              B. K. Agrawal et al., Phys. Rev. C
              {\bf 58}, 3004 (1998).
\bibitem{shlomo05} 
              S. Sholom and V. M. Kolomietz, Rep. Prog. Phys.
              {\bf 68}, 1 (2005).
\bibitem{Shetty} 
              D. V. Shetty and S. J. Yennello, Pramana
              {\bf 75}, 259 (2010).
\bibitem{kowalski} 
              S. Kowalski et al., Phys. Rev. C
              {\bf 75}, 014601 (2007).
\bibitem{DeA12} 
              J. N. De, S. K. Samaddar, and B. K. Agrawal, Phys. Lett. B
              {\bf 716}, 361 (2012).

\bibitem{Agarwal14} 
              B. K. Agrawal, J. N. De, S. K. Samaddar, M. Centelles, and X. Vi\~nas, Eur. Phys. J. A
              {\bf 50}, 19 (2014).
\bibitem{hard06} 
              P. G. Reinhard, M. Bender, W. Nazarewics, and T Vertse, Phys. Rev. C
              {\bf 73}, 014309 (2006).
\bibitem{Daniel03} 
              P. Danielewicz, Nucl. Phys. A
              {\bf 727}, 233 (2003).
\bibitem{dean02} 
              D. J. Dean, K. Langanke, and J. M. Sampaio, Phys. Rev. C
              {\bf 66}, 045802 (2002).
\bibitem{De12}
              J. N. De and S. K. Samaddar, Phys. Rev. C
              {\bf 85}, 024310 (2012).
\bibitem{Zhang14} 
              Z. W. Zhang, S. S. Bao, J. N. Hu, and H. Shen, Phys. Rev. C
              {\bf 90}, 054302 (2014).
\end{thebibliography}
\end{document}